\documentclass[apj]{emulateapj}
\usepackage{epsf}
\usepackage{amsmath}

\newcommand{\rgood}{\texttt{rgood}}
\newcommand{\igood}{\texttt{igood}}
\newcommand{\bgood}{\texttt{bgood}}

\newcommand{\Itot}{\ensuremath{I_{\rm tot}}}

\newcommand{\bootes}{Bo\"otes\ } 
\newcommand{\spitzer}{{\it Spitzer}}
\newcommand{\chandra}{{\it Chandra}}
\newcommand{\galex}{{\it GALEX}}

\newcommand{\beq}{\begin{equation}}
\newcommand{\eeq}{\end{equation}}
\newcommand{\beqa}{\begin{eqnarray}}
\newcommand{\eeqa}{\end{eqnarray}}

\shorttitle{AGES Galaxy Luminosity Function}
\shortauthors{Cool et al.}

\begin{document}
\title{The Galaxy Optical Luminosity Function from the AGN and Galaxy Evolution Survey (AGES)}
\author{
Richard J. Cool  \altaffilmark{1,2},
Daniel J. Eisenstein  \altaffilmark{3},
Christopher S. Kochanek   \altaffilmark{4} 
Michael J. I. Brown  \altaffilmark{5},
Nelson Caldwell  \altaffilmark{6},
Arjun Dey   \altaffilmark{7},
William R. Forman  \altaffilmark{6},
Ryan C. Hickox  \altaffilmark{6,8},  
Buell T. Jannuzi  \altaffilmark{7},
Christine Jones  \altaffilmark{6},
John Moustakas  \altaffilmark{10},
Stephen S. Murray  \altaffilmark{6,11}
}
  \altaffiltext{1}{The Observatories of the Carnegie Institution of Washington, 813 Santa Barbara Street, Pasadena, CA 91101}
  \altaffiltext{2}{Hubble Fellow, Carnegie-Princeton Fellow}
  \altaffiltext{3}{Steward Observatory, University of Arizona, 933 North Cherry Avenue, Tucson, AZ 85721}
  \altaffiltext{4}{Department of Astronomy, The Ohio State University, 140 West 18th Avenue, Columbus OH 43210}
  \altaffiltext{5}{School of Physics, Monash University, Clayton, Victoria 3800, Australia}
  \altaffiltext{6}{Smithsonian Astrophysical Observatory, 60 Garden Street, Cambridge, MA 02138}
  \altaffiltext{7}{National Optical Astronomy Observatory, Tucson, AZ 85726}
  \altaffiltext{8}{Department of Physics, Durham University, South Road, Durham, DH1 3LE, United Kingdom}
  \altaffiltext{9}{STFC Postdoctoral Fellow}
  \altaffiltext{10}{Center for Astrophysics and Space Science, University of California, San Diego, 9500 Gilman Drive, La Jolla, CA 92093}
  \altaffiltext{11}{Department of Physics and Astronomy, Johns Hopkins University, Baltimore, MD 21205}


\begin{abstract}
We present the galaxy optical luminosity  function for the redshift range
 $0.05<z<0.75$ from the
AGN and Galaxy Evolution Survey (AGES), a spectroscopic survey of 7.6 deg$^2$
in  the \bootes field of the NOAO Deep Wide-Field Survey.  
Our statistical sample is comprised of 12,473 galaxies with 
known redshifts down to $I=20.4$ (AB).  Our results at low
redshift are consistent with those from SDSS; at higher redshift, 
we find strong evidence for evolution in the luminosity function, 
including differential evolution between blue and red galaxies.  We find that
the luminosity density evolves as $(1+z)^{(0.54\pm0.64)}$ for red galaxies
and $(1+z)^{(1.64\pm0.39)}$ for blue galaxies. 
\end{abstract}

\section{Introduction}
The galaxy luminosity function directly quantifies the total light in 
galaxies, and its evolution characterizes the
growth of galaxies over cosmic time either through star formation or 
hierarchical assembly.  Since the first systematic galaxy redshift surveys
in the 1980s \citep{Huchra83}, the volume of the universe probed by uniform
imaging and the number of galaxies with known redshifts have grown exponentially.
With the advent of large, homogeneous, imaging and spectroscopic 
surveys of the nearby universe, such as the Two Degree Field Galaxy Redshift
Survey \citep[2dF;][]{Colless01} 
and the Sloan Digital Sky Survey \citep[SDSS;][]{York00} as well as
large-scale photometric redshift surveys such as COMBO17 \citep{Wolf03}, 
the local ($z<0.2$)
galaxy optical luminosity function is quite well constrained near $L_*$
\citep[e.g.,][]{Blanton01, Kochanek2001, Madgwick2002,
Norberg02, Blanton03LF, Bell04, Croton2005, Montero2007}.

In order to measure the evolution in the field galaxy luminosity function, one
requires measurements at several redshifts. With the advent of more powerful 
telescopes and instrumentation, a number of pencil beam surveys were 
used to quantify the galaxy luminosity density beyond $z=0.5$
\citep[e.g,][]{Lilly95, Cowie96, Brinchmann98, Lin99, 
Cohen02, Im02, deLapparent03, Cross04, Pozzetti03}.  
These pencil beam surveys, however, often 
probe volumes too small to be representative of the entire galaxy population 
(i.e. cosmic variance).  Recently, several larger area surveys, targeting
many thousands of galaxies to $z\sim1$, have allowed for more robust
statistics of the high-redshift luminosity function.  The VIMOS/VVDS
Deep Survey \citep[VVDS;][]{leFevre2004} has measured the evolution of the total
galaxy luminosity function to $z\sim1.2$ with a sample of 11,000 galaxies 
\citep{Ilbert06}.  
The DEEP2 survey \citep{Davis03} obtained redshifts for $\sim40,000$ galaxies
with DEIMOS on Keck over $\sim4$ square degrees and focused primarily on galaxies
at $z>0.7$ with one field, the Extended Groth Strip, 
used to target galaxies at all redshifts.  Comparisons between these high-redshift surveys and low-redshift
benchmarks yield our strongest current constraints on the evolution of the
galaxy luminosity function from $z=1$ to the present \citep{Willmer06, Faber07}.

While the low- and high-redshift ends of the interval between $z=0$ and $z=1$ 
have been probed with large statistical samples and volumes, intermediate-redshifts 
require an uncomfortably  large area to be spectroscopically observed to
moderate depth in order to measure the evolution of galaxy properties.  
Measurements at $z=0$ and $z=1$ can provide the overall trend with which
galaxy properties have changed over the latter half of cosmic history, but only measurements 
at intermediate-redshift characterize this evolution on finer scales.  Furthermore
surveys at $z=0$ and $z=1$ may have different systematic errors (for example 
in photometric measurements and calibration) resulting in systematic errors
when measuring evolving parameters between surveys.  Here, 
we present the evolution of the galaxy optical luminosity function 
from $0.05<z<0.75$ from the AGN and Galaxy Evolution Survey (AGES).

AGES is a spectroscopic survey of galaxies and quasars in the NOAO
Deep Wide-Field Survey \citep[NDWFS;][]{jannuzidey1999} \bootes 
 field using the Hectospec instrument
on the MMT \citep{fabricant1998, roll1998, fabricant2005}. The \bootes 
field was chosen for our redshift survey because of the  the wide array of deep multiwavelength 
photometry available in the field including ground-based optical, near-infrared,
and radio photometry as well as \spitzer, \chandra, and \galex\ imaging.
Most of these cover the full 9 deg$^2$ footprint outlined by the ground-based 
optical data.  AGES spectroscopy reached $I_{\rm AB}=20.45$ for galaxies and 
$I_{\rm AB}=21.95$ for AGN with extensions to $I_{\rm AB}=22.95$ in some regions.  The 
galaxy sample is about three magnitudes deeper than the 
SDSS MAIN galaxy sample ($r<17.7$) \citep{strauss2002} and covers about twice the area probed by 
the DEEP2 survey.  AGES is currently the largest spectroscopic 
survey of intermediate redshift field galaxies and thus provides 
an excellent sample of galaxies with which to measure the evolution of the
galaxy optical luminosity function.  

In this paper, we present a summary of the AGES galaxy sample and optical
imaging in 
\S\ref{sec:data} and give further details of the galaxy selection function in 
Appendix \ref{sec:galsec}.  Our photometry and $k$-corrections are described in 
\S\ref{sec:kcorr}. We present our luminosity function measurements in 
\S\ref{sec:gallf} including comparisons to SDSS and quantify its evolution
before concluding in \S\ref{sec:conclusions}. Throughout the paper, we use a spatially flat cosmology of 
$\Omega_m=0.3$, $\Omega_\Lambda=0.7$, and 
$H_0=100h$ km s$^{-1}$ Mpc$^{-1}$.  We use AB magnitudes for all bands 
\citep{oke1974}, although the photometric catalogs
from the NDWFS
use Vega magnitudes\footnote[1]{ We adopt AB corrections of:
$B_{W,AB}=B_{W,{\rm Vega}}$, 
$R_{AB}=R_{{\rm Vega}}+0.21$, 
$I_{AB}=I_{{\rm Vega}}+0.45$}.

\section{Optical Imaging and AGES Sample}
\label{sec:data}

\subsection{Optical Imaging}
We use the deep optical ($B_WRI$) photometry from the 9.3 deg$^2$ 
\bootes field provided by the third data release from the NOAO 
Deep Wide-Field Survey \citep{jannuzidey1999}.  
A full description of the observing 
strategy and data reduction is presented elsewhere (Jannuzi et al., {\it in prep}; Dey et al., {\it in prep}) and the data can be obtained publicly
from the
NOAO Science Archive\footnote[2]{http://www.archive.noao.edu/ndwfs \\ http://www.noao.edu/noao/noaodeep}. 
The NDWFS catalogs reach $B_{w,\mbox{\scriptsize AB}}\sim26.5$, 
$R_{\mbox{\scriptsize AB}}\sim25.5$, $I_{\mbox{\scriptsize AB}}
\sim25.3$, and $K_{s,\mbox{\scriptsize AB}} \sim23.2$ at 50\% completeness for point and 
are more than 85\% complete for galaxies of typical sizes and shapes to $I_{\mbox{\scriptsize AB}}=23.7$ 
\citep{Brown07}. Here,
we utilize photometry from the DR3 release of NDWFS imaging.
When performing $k$-corrections of AGES sample galaxies, we augment the
NDWFS imaging with 8.5 deg$^2$ (covering 7.7 deg$^2$ of the NDWFS footprint)
of $z'$-band imaging from the zBootes survey \citep{Cool2007}.  The 
typical $3\sigma$ depth of these catalogs is 22.5 mag for point sources
in a 5 arcsecond diameter aperture.  Full details of the data reduction and 
a full release of the $z'$-band imaging catalogs can be found in 
\citet{Cool2007} and the NOAO Science Archive\footnote[3]{http://archive.noao.edu/nsa/zbootes.html}. 

\subsection{AGES}
AGES used the Hectospec instrument on the MMT to survey 9 deg$^2$ of the 
NDWFS \bootes field.  Full details of the survey will be given in \citet{Kochanek2011};
here, we describe only the aspects relevant to the galaxy survey.
The final AGES galaxy sample was selected from the NDWFS optical imaging 
catalogs to $I_{\rm AB}<20.45$.  In addition, galaxies must satisfy image quality
cuts in the $I$ band and at least one of the $B_W$ or $R$ bands. 
As the imaging data is several magnitudes deeper than the spectroscopic sample, in 
order for a galaxy to fail the $B_w$ or $R$ detection thresholds due to 
galaxy color would imply a galaxy with colors much more extreme than found in existing
surveys such as the SDSS.  Thus, the requirement to include all three bands primarily
limits the total available survey area as it excludes regions where one of the imaging data sets
is missing.  In 
order to be included in the sample we utilize for luminosity function 
measurements, we require a good quality detection in all three $B_W$, 
$R$, and $I$ bands in order to ensure robust and uniform $k$-corrections. 
The NDWFS imaging is considerably deeper than the spectroscopic flux limit,
thus requiring detections in all three bands does not bias our sample
even for galaxies with peculiar optical colors. Finally, we require that
galaxy targets were detected as a non-point source in at least one of the 
$B_W$, $R$, or $I$ bands. The requirement that the galaxy be extended in 
at least one of the three imaging bands places a physical size requirement into the sample, but this
limit is not physically interesting for galaxies at the redshift and luminosities probed by our sample.  
For example, for a galaxy at $z=0.5$ to fall below $r=1.2''$, it would correspond to a physical size of
$r\sim7.4$ kpc.  At the luminosity depth of the survey at $z=0.5$, galaxies small enough to be unresolved
by NDWFS imaging and yet pass the AGES flux limit would represent a currently unknown population of 
extremely luminous ultra-compact galaxies and thus we do not consider this a likely source of
incompleteness in our sample.

AGES employed a complex set of sparse sampling criteria  
where galaxies with $I<20.45$ which were also bright at other wavelengths (including 
in \spitzer, \chandra, or \galex\ imaging) were more likely to be observed.
However, the sampling fraction in each galaxy subsample is known and hence
we can correct for this sampling function when constructing our luminosity
function measurements. The lowest sampling rate (for galaxies with 
$18.95<I<20.45$ that failed all other targeting cuts) was 20\%. By
weighting the galaxies by the inverse of the sampling rate, we can 
restore a statistically uniform sample with $I<20.45$.  Further details 
about our selection function can be found in Appendix \ref{sec:galsec}.

We observed our targets at the MMT 6.5~m telescope over three years, 
2004--2006.  The time allocation in 2006 was aimed at fainter AGNs; only a
few remaining galaxies were targeted, which we include here. The
Hectospec instrument has a 1 degree diameter field of view patrolled by 300 
robotically positioned fibers.  An atmospheric dispersion corrector (ADC)
ensures that light losses due chromatic effects are minimized in the $1\farcs5$ diameter fibers.
The fibers feed into a single spectrograph with a 270/mm grating which
yields 6 \AA\ FWHM spectra.  The data were reduced with two separate 
spectroscopic pipelines, described in \citet{Kochanek2011}.  In 2004, some observations
were done with the ADC functioning improperly.  While the loss of light
on these observations greatly impacts the spectrophotometry of the resulting
spectra, we were still able to obtain redshifts for the vast majority of the observed
galaxies; galaxies which failed to generate redshifts with these observations
were re-observed in later years. 

In 2004, we covered most of the \bootes field with 15 pointings, each with 3
fiber configurations. These 15 pointings cover 7.6~deg$^2$ and are taken
to define the galaxy survey region.  Although we observed outside of this 
primary region in 2005 in order to maximize our AGN coverage, 
we downweighted galaxy targets outside of the 2004 region and thus exclude
objects outside that region in our analysis. In 2005 and 2006, we covered
the field with 63 configurations.  With a total of 108 configurations, 
plus the overlaps between the circular fields of view, each target galaxy
had many geometrical opportunities to be included in a fiber configuration.  In detail,
the target selection in 2004 was more restrictive than in 2005.  Hence, some 
objects were available to be observed both years while others were only
available during the 2005 and 2006 observations; we account for this effect
in the survey selection function which is described in detail in Appendix \ref{sec:galsec}.

\begin{figure}[b]
\plotone{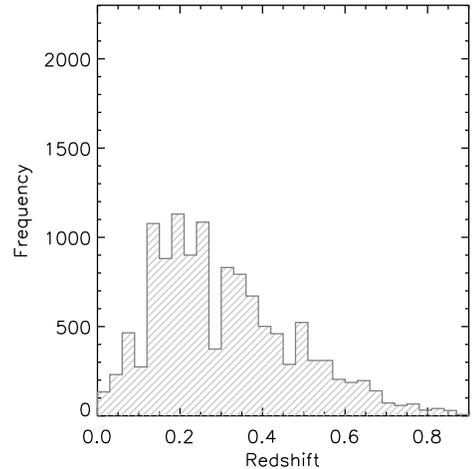}
\caption{\label{fig:zhist} Redshift histogram of primary sample galaxies from AGES (grey).  
The unshaded histogram shows the re-weighted sample after the 
affects of our a priori sparse sampling have been removed from the survey. 
 From this histogram alone, one can see the effects of large-scale structure, 
even on our 7.6 deg$^2$ field.}
\end{figure}

\begin{figure}
\plotone{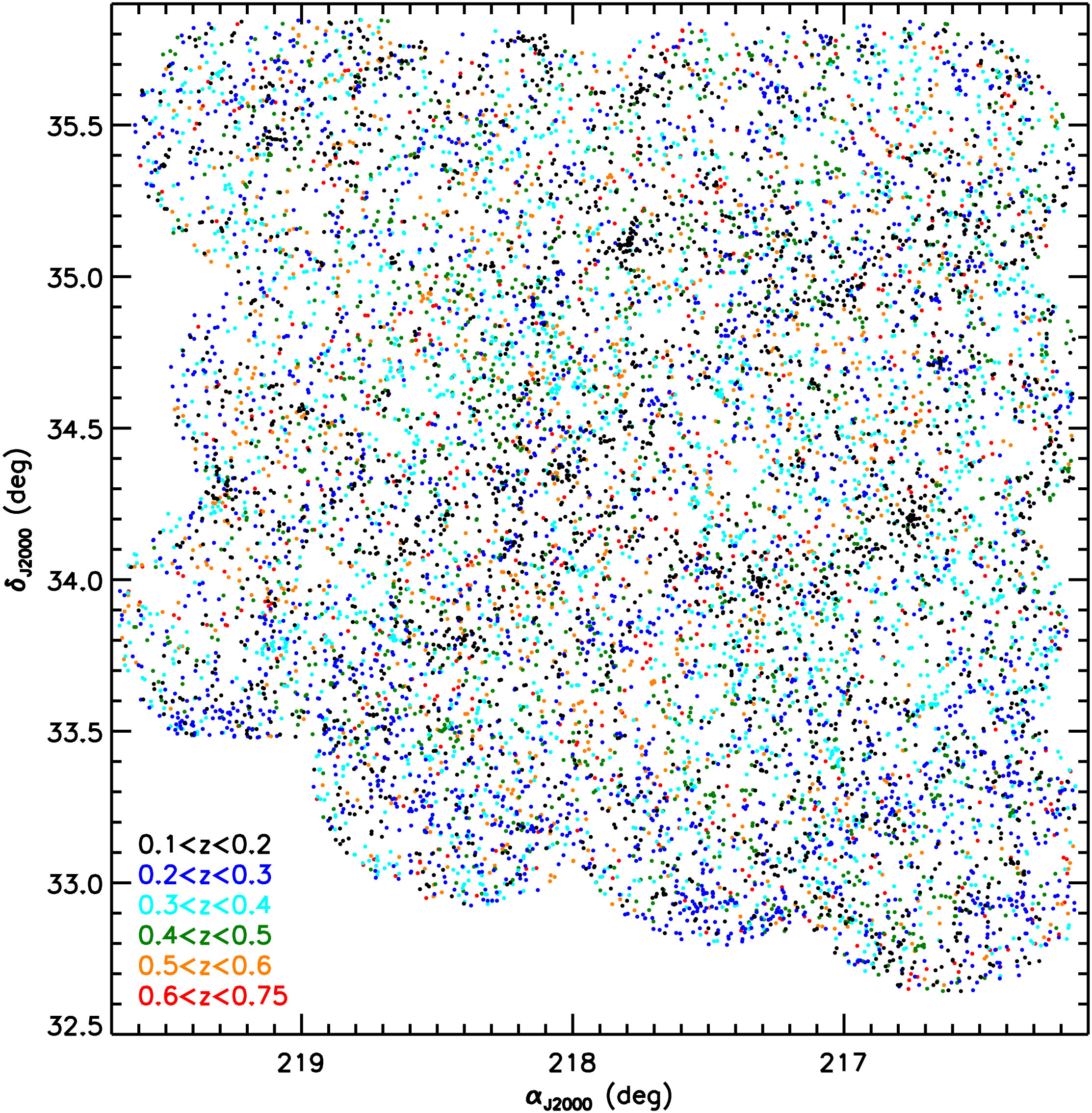}
\caption{\label{fig:window} Angular distribution of AGES spectroscopic galaxies. The
points show the spatial distribution of galaxies at $0.1<z<0.75$.  At all redshifts, large scale structure is apparent. The circular
boundaries are defined by the circular field of view of the Hectospec instrument.  While some spectroscopy exists outside this footprint, we
only include galaxies in the statistically complete main survey field (AGES fields 1-15) in this figure and in our analysis.}
\end{figure}

Because of the flexibility of the robotic fiber positioners and the monthly queue campaigns, AGES was observed with rolling target acceptance.  After each
observing run, galaxies which failed to yield a redshift were placed back into 
the queue for subsequent runs.  As a result, AGES has a very high 
spectroscopic success rate. Figure \ref{fig:zhist} shows the final
distribution of redshifts from AGES spectroscopy in the shaded region.  The 
full reweighted sample (using the procedure described in Appendix \ref{sec:galsec})
is shown by the unfilled histogram.  Over our 7.6 deg$^2$ field, the presence of 
large scale structure is apparent. Figure \ref{fig:window} shows the two-dimensional
distribution of AGES sources color-coded by measured redshift. 
The circular regions in Figure \ref{fig:window} arise from the circular 
field of view of the Hectospec instrument. 

We give full details of the selection function in Appendix \ref{sec:galsec}. 
In brief, the parent galaxy sample has 26,033 galaxies brighter than $I_{AB}=20.45$. We
estimate this photometric sample to be 4\% incomplete.  AGES observed 
approximately 50\% of this parent sample, 12,473 galaxies. Nearly all 
of the difference is due to our {\it a priori} sparse sampling which can be
corrected exactly with our known targeting rates. The sample has a
4.3\% incompleteness in assigning fibers to targets and a 2.1\% 
incompleteness in measuring a reliable redshift from an 
assigned fiber. The Appendix describes our modeling of these
incompletenesses, but the main point is that AGES is highly complete and that
the error in our incompleteness corrections are smaller
than our statistical uncertainties.

\section{Photometry and $K$-Corrections}
\label{sec:kcorr}

\subsection{$k$-corrections}
We construct our luminosity functions from the $I$-band NDWFS photometry using the 
\texttt{Sextractor} \citep{sextractor} AUTO (Kron-like \citep{Kron1980}) magnitude.  
The $I$-band photometry is contaminated in certain regions
from the low-surface-brightness wings of bright stars which bias the AUTO magnitudes brighter than reality.
We address this by constructing a surrogate $I$-band total magnitude, $I_R$ from the $R$-band AUTO magnitude
plus the $I-R$ color measured in a $6''$ aperture (as the aperture colors are much less sensitive to the low
surface brightness tails which affects the $R$-band imaging much less than the $I$-band). 
We then compare $I_R$ to the $I$-band AUTO 
magnitude and compute $I_{\rm tot}$. In the case of two significantly different values between $I$ and $I_R$, 
$I_{\rm tot}$ is the fainter of the two magnitudes. Otherwise, $I_{\rm tot}$ is the average of the two.  Explicitly,
we compute 
\begin{equation}
f= \exp[-(I_{\rm AUTO}-I_R)^2/0.2^2]
\end{equation}
and use this to linearly combine the average flux of the pair and the smaller flux of the pair
\begin{equation}
I_{\rm tot} = \frac{I_{\rm AUTO}+I_{\rm R}}{2}f+(1-f){\rm max}(I_{\rm AUTO}, I_{\rm R}).
\end{equation}
About
10\% (5\%) of the galaxies have a correction of more than 0.1 (0.5) magnitudes. Most of the galaxies have
a tight correlation between $I_{\rm tot}$ and $I_{\rm AUTO}$ with an rms of 0.02 mag and 
$\langle I_{\rm tot}-I_{\rm AUTO} \rangle=0.005$.  Because of this slight scatter, we cut our statistical
sample of galaxies at $\Itot<20.40$; this excludes 2\% of the galaxies.

When computing $k$-corrections, we estimate the 
$B_W$, $R$, and $z'$ magnitudes by adding $4''$ aperture  $B_W-I$ and $R-I$ and $I-z$ colors to $\Itot$.
In other words, the SED (colors) of each galaxy is determined by  the $4''$ aperture colors while the total amplitude 
is set by the $\Itot$ magnitude.  We find only 1\% shifts in the colors of galaxies when we use PSF-matched images
to measure aperture colors. Such shifts do not affect any of our results. 

All $k$-corrections are computed using \texttt{kcorrect v4\_2} \citep{Blanton2007}. This procedure uses a linear combination of 
galaxy spectral templates with the measured photometry and redshifts to construct a best-fit spectral energy distribution (SED) for
each galaxy in our sample. These best fitting SEDs are then used to predict the luminosity and colors of each galaxy as a function of redshift.
In order to minimize the 
effects of the $k$-corrections on our final results, we shift to bands that minimize the change in rest-frame
wavelength between observed and rest wavelength at a typical AGES redshift.  We use the $I$ band to construct the
$^{0.1}r$ luminosity.  Throughout the paper we utilize a $^{0.1}ugriz$ filter system.  In this system
the notation $^{0.1}r$ denotes the luminosity in the SDSS $r$ band \citep{fukugita1996} shifted blueward by $z=0.1$.  This is least sensitive
to the SED model at $z=0.42$ where the observed $I$-band matches the rest wavelength of $^{0.1}r$ .
This choice also makes it easy to compare to the SDSS luminosity function \citep{Blanton03LF}.  
In order to compare our results to those in the literature, we also use the $B_W$ band
to construct the $B$-band luminosity for each target.

\subsection{Red and Blue Galaxies}

\begin{figure}
\plotone{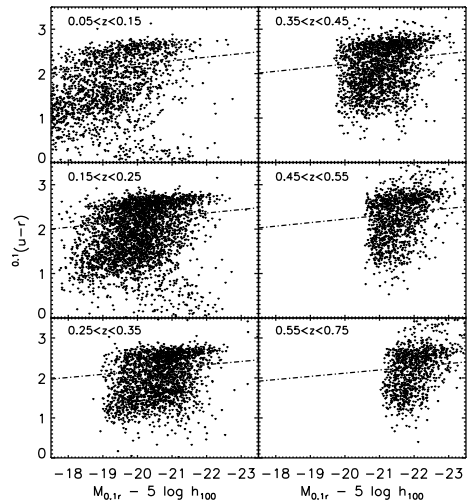}
\caption{\label{fig:cmr} Restframe $^{0.1}$(u--r) versus $M_{^{0.1}r}$ color-magnitude diagram from AGES. 
The bimodality of galaxy optical colors is apparent in each redshift bin. 
 The dot-dashed line shows the criteria used to separate red and blue galaxies as defined in the text.}
\end{figure}

In order to probe the evolution of red and blue galaxies separately, we first need to select
each type of galaxy as a function of redshift in the AGES sample. As galaxies, especially on the red sequence, 
have undergone substantial passive evolution since $z=1$, a single rest-frame color cut will not lead to a 
a homogeneous sample of red and blue galaxies over a wide range of redshifts.  Here, we solve for the evolution
in the red-sequence zeropoint empirically and use that cut when defining red and blue galaxies in the sample.
We first construct the luminosity-dependent statistic
\begin{equation}
A=^{0.1}u-^{0.1}r + 0.08 (M_{^{0.1}r}+20).
\end{equation} 
We then iteratively find the median value of $A$ for galaxies with $A>A_{{\rm med},{n-1}}-0.3$ where $A_{{\rm med},{n-1}}$
is the previous median
value; the cutoff value of $A=0.3$ was chosen to best localize the minimum of the galaxy number distribution in color
at $z=0.1$.  We use only galaxies with $M_{^{0.1}r}<-20$ in this procedure.  These medians converge to the median color
of the red sequence at $M_{^{0.1}r}=-20$. We perform these medians in redshift bins of $\Delta z=0.1$ and find values of 
$A_{{\rm med}}=2.74, 2.69, 2.64, 2.60, 2.56, 2.52$ for bins centered at redshifts 0.1, 0.2, 0.3, 0.4, 0.5, and 0.65 respectively.  
We then define our red and blue galaxy samples by linearly interpolating the $A$ parameter to the redshift of each galaxy;
galaxies which have $A_{\rm gal} < A_{\rm median}(z) -0.3 $ are classified as blue while those redward of that limit are 
classified as red.
Figure \ref{fig:cmr} shows the $^{0.1}(u-r)$ versus $M_{^{0.1}r}$ 
color-magnitude relation in six redshift slices from the AGES sample. The bimodality in galaxy colors is clearly seen in each 
slice.  The color cut used to differentiate between red and blue galaxies in each slice is also shown.

\subsection{SDSS Comparison Sample}

At low redshift, the AGES luminosity function is significantly impacted by large-scale structure and the 
limited volume we probe.  We therefore use the NYU Value-Added Galaxy Catalog (VAGC)
 to construct a comparison sample of 571,909 galaxies from the SDSS \citep{Blanton_VAGC} based on the SDSS DR7. 
   From this sample, we extract all galaxies with $0.01<z<0.15$, which we refer to as the SDSS sample.

\section{The Galaxy Optical Luminosity Function}
\label{sec:gallf}

Based on the galaxy selection function described in Appendix \ref{sec:galsec}, we can reconstruct a statistical sample
of $\Itot<20.40$ galaxies.  We calculate luminosity functions for these galaxies using two methods, the $1/V_{\rm max}$ method 
and using parametric maximum likelihood models.  

\subsection{The $1/V_{max}$ Method}

The $1/V_{\rm{max}}$ method is one of the more simple and intuitive forms for deriving the luminosity function.  
The $1/V_{\rm{max}}$ method benefits from being calculated without the need for an a priori parametric form
of the luminosity function.  In this paper we follow the techniques described in
\citet{Eales1993}, \citet{Lilly95}, \citet{Ellis1996}, \citet{Takeuchi2000}, and \citet{Willmer06}.  For each
galaxy in our sample, we first calculate $z_{\rm{max}}$ and $z_{\rm{min}}$, the full range of redshift for which the galaxy
may have been selected through our direct $\Itot<20.4$ 
cut including effects of the $k$-correction
and luminosity distance. 
 We then calculate the maximum volume each galaxy could 
have occupied and been included when considering all galaxies in a given redshift range set by $z_{\rm{lower}}$ and 
$z_{\rm{upper}}$:
\begin{equation}
V_{\rm{max}}(i) = \int_\Omega\int^{z_{\rm{max},i}}_{z_{\rm{min}, i}} \frac{d^2V}{d\Omega dz} dz d\Omega . 
\end{equation}
Here, $z$ is the redshift and $\Omega=7.6\,\rm{deg}^2$  is the solid angle covered by the survey.  The limits of the inner integral
are given by 
\begin{equation}
z_{\rm{max},i} = {\rm min}\{ z_{\rm{max}}, z_{upper}\} 
\end{equation}
\begin{equation}
z_{\rm{min},i} = {\rm max}\{ z_{\rm{min}}, z_{lower}\}.
\end{equation}
Once we have the $V_{max}$ values for each galaxy in a redshift slice, we can calculate the integral luminosity 
function in a given magnitude range with $M_{\rm bright}<M<M_{\rm faint}$ using :
\begin{equation}
\phi(M) \Delta M= \sum_{i=1}^{N_{\rm  gal}}\frac{w_i}{V_{\rm{max}}(i)}.
\end{equation}
Here, the $w_i$ are the statistical weights described in Appendix \ref{sec:galsec} used to correct our sample into a
full $\Itot<20.4$ flux-limited sample and $\phi$ is the luminosity function.
The error for estimates in the $1/V_{\rm{max}}$ method are determined 
by:
\begin{equation}
\sigma_\phi = \sqrt{\sum_i \frac{w_i}{V_{\rm{max}}(i)^2}}
\end{equation}

\subsection{Parametric Maximum Likelihood Methods}

We also use a parametric maximum-likelihood fit to a Schechter function using the
STY estimator \citep{Sandage1979, Efstathiou1988}.  We utilize the standard \citet{Schechter1976} function
of the form:
\begin{equation}
\phi(L)dL = \phi_*(L/L_*)^\alpha e^{-L/L_*}dL .
\end{equation}

Here, $L_*$ characterizes the break in the luminosity function, $\alpha$ sets the faint-end slope, and
$\phi_*$ is the normalization.  Re-writing this in magnitudes we obtain
\begin{equation}
\phi(M) dM = 0.4 \mbox{ln}(10) \phi_* 10^{-0.4(M-M_*)(\alpha+1)}\exp\left(-10^{-0.4(M-M_*)}\right)dM.
\end{equation}
The maximum likelihood method tests the free parameters of the parameterization ($M_*$ and $\alpha$ here, but not
$\phi_*$) by calculating the probability of observing a galaxy with luminosity $M$, given its observed redshift, $z$, and
the survey selection function $s(M,z)$.   Explicitly, the likelihood for galaxy, $j$, is 
\begin{equation}
\label{eqn:pz}
p(M_j,z_j) = \frac{ \phi(M_j)s(M_j,z_j)}{\int \phi(M) s(M,z) dM}
\end{equation}
where the integral is over the full luminosity range of the survey and the selection function, $s(M,z)$, 
includes information about the sample flux-limit.  The corresponding quantity for the full sample is the 
product of the likelihood over all galaxies.  Finally, we maximize the quantity
\begin{equation}
\mathcal{L} = \sum_j {\rm log}\ p(M_j|z_j)
\end{equation}
as a function of the $M_*$ and $\alpha$ parameters.

Due to the ratio in equation (\ref{eqn:pz}), the normalization, $\phi_*$ is unconstrained by the STY method.  
We use the minimum variance estimator of \citet{DavisHuchra82} to perform the normalization
\begin{equation}
\label{eqn:n}
n = \frac{\sum_j f(z_j)}{\int dV \eta(z) f(z)}
\end{equation}
where $n$ is mean number density of galaxies in the sample,
\begin{equation}
f(z) = \frac{w_i}{1+nJ_3 \eta(z)}
\end{equation}
is the weight for each galaxy
 and the selection function is given by 
\begin{equation}
\eta(z) = \frac{\int_{max(L_{\rm min}(z), L_{\rm faint})}^{min(L_{\rm max}(z), L_{\rm bright})}dL\phi(L)}{\int_{L_{\rm faint}}^{L_{\rm bright}}\phi(L)dL}.
\end{equation}
 The luminosity function normalization
is given by 
\begin{equation}
\phi_* = \frac{n}{\int^{M_{\rm faint}}_{M_{\rm bright}} \phi(m) dM}.
\end{equation}
The contribution of galaxy clustering to the number density is 
accounted for using the second moment, $J_3 = \int_0^r r'^2 \xi(r') dr'$, 
 of the two-point correlation function $\xi(r)$.
Since $n$ appears in the weight, $f(z)$, we determine it iteratively using 
\begin{equation}
n= \frac{1}{V}\sum_j\frac{1}{\eta(z_j)}
\end{equation}
as our initial guess. 
Finally, when calculating the luminosity density, we integrate the best-fitting
luminosity function
\begin{equation}
j(z) = \int L\phi(L) dL = L^*\phi^*\Gamma(\alpha+2)
\end{equation}
where $M_{\odot,r^{0.1}}=4.76, M_{\odot, B}=5.48$, and $\Gamma$ is the Gamma function.
This form of integration to measure the luminosity density does not depend on the
limiting magnitude of the survey but does depend on the form of the luminosity function
in order to extrapolate to the entire population.  Since the majority of the luminosity
is located at $M^*$ for a population following a Schechter function, only the highest
redshift bins in this survey extrapolate more than 50\% of the luminosity density when
calculating the total $j$ compared to the luminosity range probed by the survey; at the 
lowest redshifts considered, we probe 80-95\% of the total galaxy luminosity density.

\subsection{Luminosity Function Results}

\begin{figure}
\plotone{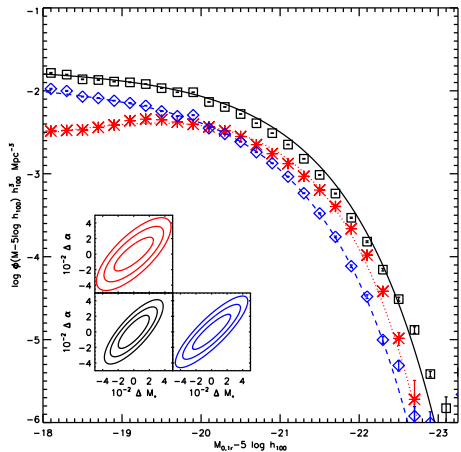}
\caption{\label{fig:sdsslf} SDSS galaxy sample $^{0.1}r$ luminosity functions for all (black squares), 
blue (blue diamonds), and red (red asterisks) galaxies
as defined in the text.  As AGES is limited by large-scale structure in the field at the lowest redshifts, we use these SDSS luminosity functions 
as an anchor for the evolution we measure based on AGES luminosity functions at higher redshifts.  Furthermore, we
fit the full shape of the SDSS luminosity functions using the maximum likelihood technique (fits shown with dot-dashed lines)
to fit the faint-end slope, $\alpha$,  of the galaxy optical luminosity function at $z=0.1$.  Since our AGES luminosity functions
do not extend significantly fainter than $L_*$ at $z>0.3$, we fix the $\alpha$ values used when fitting AGES LFs to those 
determined from SDSS. Note that these agree with the estimates from the lowest redshift AGES bin but have smaller uncertainties.
The inset illustrates the covariance associated with the $M_*$ and $\alpha$ measurements based on these luminosity functions. }
\end{figure}

Figure \ref{fig:sdsslf} shows the resulting optical luminosity functions in the
$^{0.1}r$ band for SDSS galaxies at $0.01<z<0.15$.    
The points are the 1/$V_{\rm max}$ result and the lines 
are the best fitting STY luminosity functions.
 We used this SDSS luminosity function to fit the value
of the Schechter $\alpha$ parameter when fitting the AGES data, as the 
AGES luminosity function in the  lowest redshift bin are limited by the 
small volume probed.  We restrict our fitting
of the SDSS data to galaxies between $-23 < M_{^{0.1}r} < -18$. 
This range was chosen to mirror the luminosities probed by our AGES
galaxy sample.  In order to test the effects on this range, we also fit the SDSS
data
with faint limits to $M_{^{0.1}r}<-17$ and $M_{^{0.1}r}<-17.5$; the range used in the fitting
has no strong effect on the derived $\alpha$ parameters.  Using 
STY determined luminosity function parameters (shown as lines
in Figure \ref{fig:sdsslf}), we find $\alpha=-1.05\pm0.01$ for the full galaxy 
sample, $\alpha=-1.15\pm0.02$ for the blue galaxy sample, and $\alpha=-0.50\pm0.02$ for the
red galaxy sample.  Throughout the rest of our discussion, we hold
these $\alpha$ values fixed for each survey when fitting the AGES data.
For comparison, the lowest redshift bin in AGES yields values of 
$\alpha=-1.06\pm0.03$ for all galaxies, $\alpha=-0.46\pm0.09$ for red galaxies, and
$\alpha=-1.15\pm0.04$ for blue galaxies which is in excellent agreement
with the SDSS values but with larger uncertainties. 
In order to test the overall normalization of our luminosity function measurements, 
we calculate the number counts as predicted from the final derived parameters and compare these
to the AGES target number density.  We predict $3310.1\pm118$, $1403.1\pm80.1$, $1850.1\pm118$ galaxies per square degree for
all, red, and blue galaxies respectively for galaxies with $I<20.4$ and $0.05<z<0.75$; the error term is dominated by the range of $k$-corrections
associated with each population.  The AGES number counts for galaxies in this magnitude and redshift range
are 3269.61, 1420.8, and 1875.13 for all, red, and blue galaxies.

\begin{figure}
\plotone{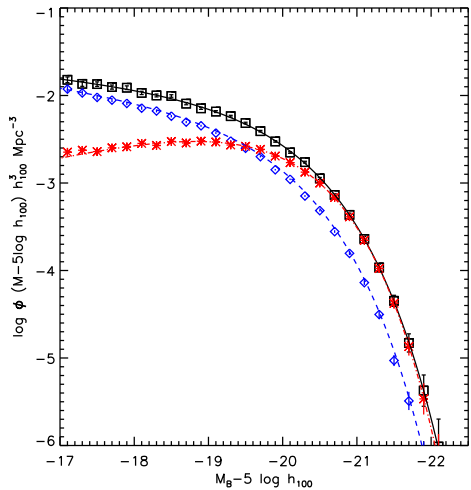}
\caption{\label{fig:sdssBlf} Same as Figure \ref{fig:sdsslf} but showing the B-band luminosity function for SDSS galaxies at 
$z=0.1$.  Again, we fit the SDSS luminosity functions to determine the local $\alpha$ value in the $B$-band and then 
assume $\alpha$ is fixed in redshift.}
\end{figure}

As the luminosity function has historically been measured in the $B$-band,
we also $k$-correct the SDSS galaxy photometry to the restframe ($z=0$) $B$-band and
measure the galaxy $B$-band luminosity function from SDSS for comparison.  
Figure \ref{fig:sdssBlf} shows the $B$-band luminosity functions for 
 SDSS galaxies at $0.01<z<0.15$.  Again, we use the $\alpha$ values derived
from these low-redshift fits to fix the values used when fitting AGES luminosity
functions at higher redshift.  In the $B$-band we find that $\alpha=-0.99\pm0.02$ for
the full SDSS sample, $\alpha=-1.10\pm0.03$ for blue galaxies and $\alpha=-0.45\pm0.02$ for 
red galaxies. 

\begin{figure}
\plotone{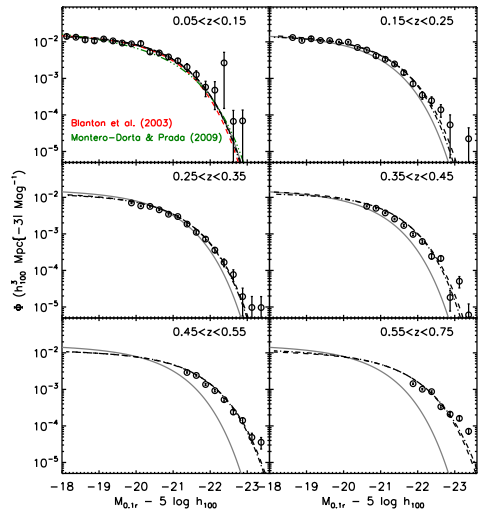}

\caption{\label{fig:agesr} AGES $^{0.1}r$ luminosity function for all galaxies at $0.05<z<0.75$.  Each panel shows the
optical luminosity function in bins of increasing redshift with the redshift range listed in the upper right corner of the panel. 
The dashed line shows the best fitting
STY parameterization while the data points with error bars show the
$1/V_{\rm max}$ measurements. The 
data points show the $1/V_{max}$ method determination of the luminosity function and the grey line shows the SDSS $z=0.1$ LF
for all galaxies.  
Evolution is clearly detected; galaxies in the past were, on average, more luminous than they are at the present epoch. 
The dot-dashed lines show the sum of the blue and red galaxy luminosity functions shown in Figures \ref{fig:agesr_blue}
and \ref{fig:agesr_red}.  The total luminosity density from AGES agrees well with the total of the red and blue populations.  Additionally,
we plot the $z=0.1$ SDSS luminosity functions from \citet{Blanton03LF} and \citet{Montero2007} for comparison.}
\end{figure}

\begin{figure}

\plotone{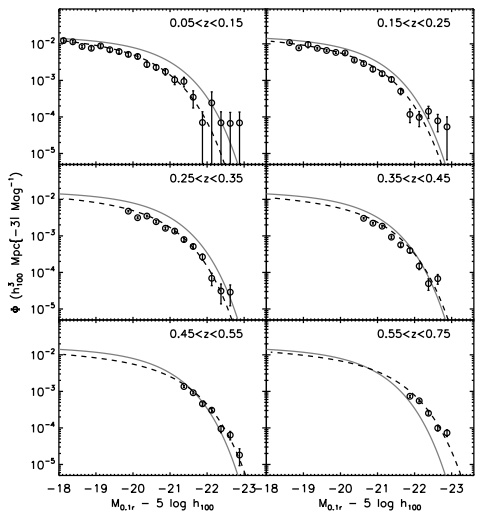}
\caption{\label{fig:agesr_blue} Same as Figure \ref{fig:agesr}, but only including blue galaxies as defined by 
the empirical red/blue classification described in the text and shown in Figure \ref{fig:cmr}.   
For reference, the $z=0.1$ LF for all galaxies determined from 
SDSS data is shown in grey. }
\end{figure}

\begin{figure}

\plotone{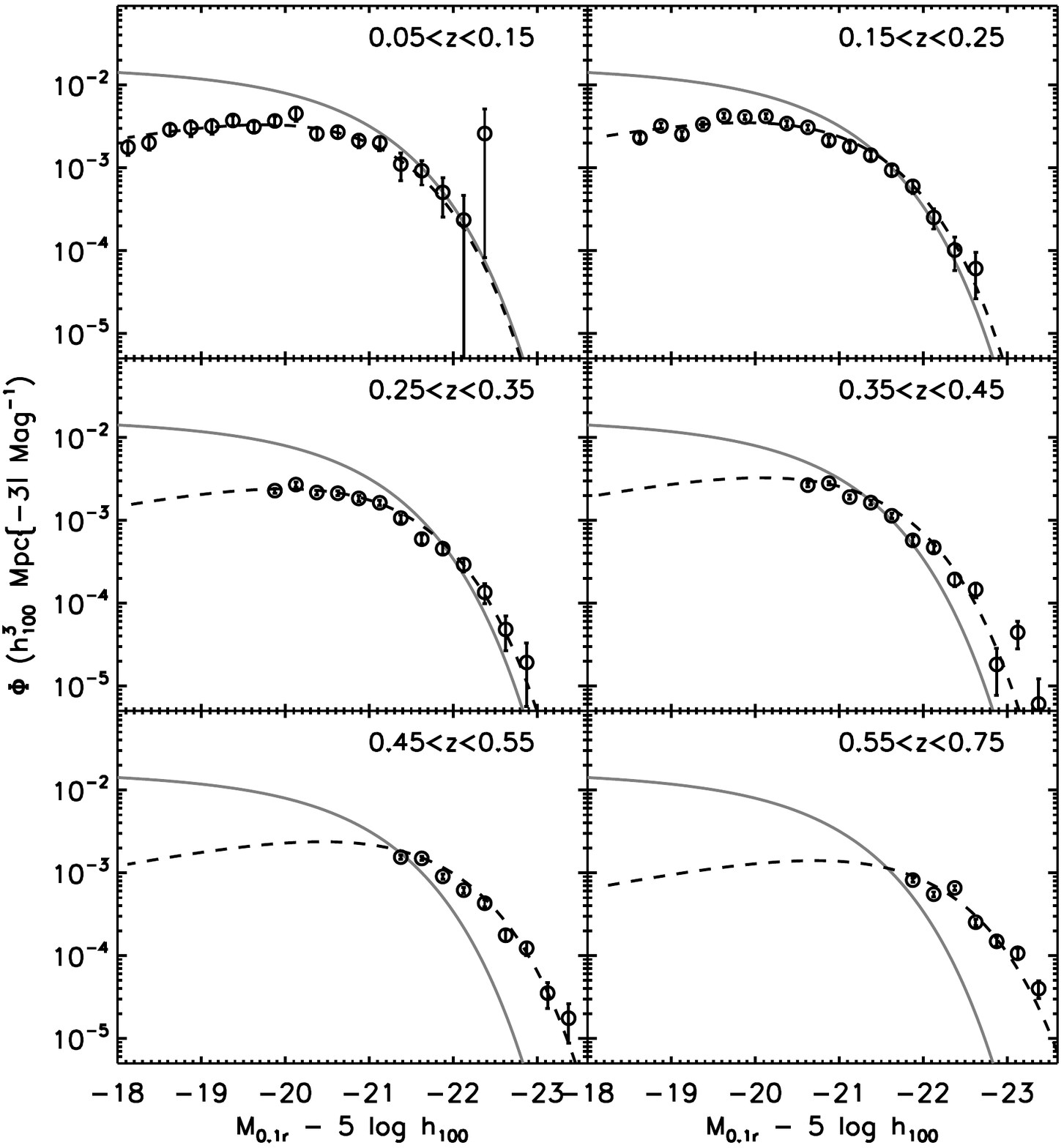}
\caption{\label{fig:agesr_red} Same as Figures \ref{fig:agesr} and \ref{fig:agesr_blue}, but now for red galaxies.  Again, 
the grey line shows the LF for all galaxies at $z=0.1$ from SDSS.}
\end{figure}

We show the $^{0.1}r$-band AGES galaxy optical luminosity functions  based on the
$1/V_{max}$ method in Figures 
\ref{fig:agesr}-\ref{fig:agesr_red} and 
list our $1/V_{\rm max}$ measurements in Tables \ref{tab:ages_r_All_lf}-\ref{tab:ages_r_Red_lf}.   
In each figure, the grey line shows the $z=0.1$ luminosity
function of all SDSS galaxies.  The data points show the 
$1/V_{\rm max}$ luminosity function measurements and the solid dashed lines show 
the STY determined parameterization. 
As an added check on our luminosity function calculation, we 
further plot the summed red galaxy and blue galaxy luminosity functions to compare the with 
total galaxy luminosity function in Figure \ref{fig:agesr}; the summed
luminosity function agrees well with the total galaxy luminosity function.
 The typical luminosity of galaxies increases with redshift. 
 That is, the stellar populations
of galaxies of all colors has, on average, faded from $z=0.75$.
The best fitting parameters determined from our STY fits
are listed in Table \ref{tab:lffits}.

We estimate uncertainties for our fitted parameters using a two-part strategy.  First, 
to include the effects of small-scale structure and shot noise, we
use a jackknife method.  We split the sample
into 15 roughly equal-area subregions on the sky (the 15 AGES fields) and repeat our fits excluding
one area at a time.  The rms variation between N such samples, multiplied
by $\sqrt{N-1}$, is an estimate of the uncertainty in each parameter.   In 
this test, we calculate the luminosity density for each subsample before the 
variance is computed in order to account for the covariance between $M_*$ and 
$\alpha$. 

These jackknife error estimates would be correct if the subregions were 
statistically independent from each other.  However, it does not account for
the correlations between subregions due to large-scale structure, most notably
 on scales larger than the survey region as a whole.  This
large-scale structure contribution can be calculated from the two-point 
correlation function. We compute the large-scale structure
contribution by averaging the correlation function across all pairs in a given 
redshift shell within our survey volume.  This gives the variance 
of the fractional over density in the survey region.  As the
AGES survey is reasonably large, we use the linear-regime power spectrum 
from the WMAP-3 cosmology \citep{Spergel2007} to generate the correlation function.  The
averaging over pairs is accelerated by projecting the partial correlation function
to the angular correction function using the Limber approximation 
and then performing a Monte Carlo angular integration using
a set of points randomly distributed in the survey area.  The results 
differ only modestly from the rms density fluctuations in a
sphere whose volume equals that of the
survey in the given redshift bin.  

\begin{figure}
\plotone{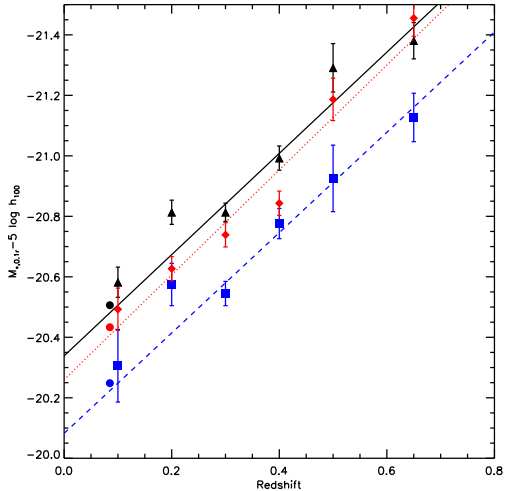}
\caption{\label{fig:rmstar} Best fitting $M_*$ in the $^{0.1}r$ band based on STY
fits to the AGES lgalaxy sample.   The empty symbols show the AGES data for all galaxies (black triangles), 
red galaxies (red diamonds), and blue galaxies (blue squares).  The filled circles show the best fitting SDSS 
values (the SDSS measurement is at $z=0.1$, but has been shifted  to slightly lower redshift for plotting clarity).  
All galaxies in the survey evolve similarly although the red galaxies fade slightly faster with redshift than the
full galaxy sample or blue galaxies.  We find that $M_*$ for the full sample fades by $1.67\pm0.07$ mag per unit redshift, 
the blue sample fades by $1.66\pm0.09$ mag per unit redshift
while the red galaxies fade by $1.73\pm0.07$ mag per unit redshift. The lines show the best fitting 
evolution for all (solid), red (dotted), and blue (dashed) galaxies.}
\end{figure}

\begin{figure}
\plotone{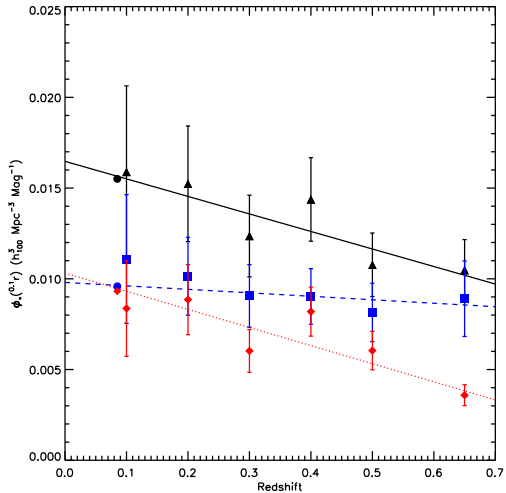}
\caption{\label{fig:rphistar} Best fitting $\phi_*$ values in the $^{0.1}r$ band determined from our STY analysis. 
 As in Figure \ref{fig:rmstar}, the full galaxy sample is plotted with black triangles, 
the red galaxies with red diamonds, and the blue galaxies with blue squares.  The best fitting
evolution is shown with solid, dotted, and dashed lines, respectively.  We find 
that the number density of all galaxies has increased since $z\sim0.7$ with the red galaxies showing the most dramatic increase in number density.  The
fits are constrained to go through the SDSS measurement at $z=0.1$ (filled circles),
 as the lowest-redshift bin from AGES is highly susceptible to large scale structure
due to its relatively small area of the survey. When fitting the data, we use a model of the form $\phi_*\propto 1+Pz$. 
 We find that the trend in the full population ($P=-0.59\pm0.14$) is driven primarily by a sharp decline in the 
red population ($P=-0.95\pm0.10$) while the blue galaxies show much less evolution ($P=-0.38\pm0.21$).}
\end{figure}

\begin{figure}
\plotone{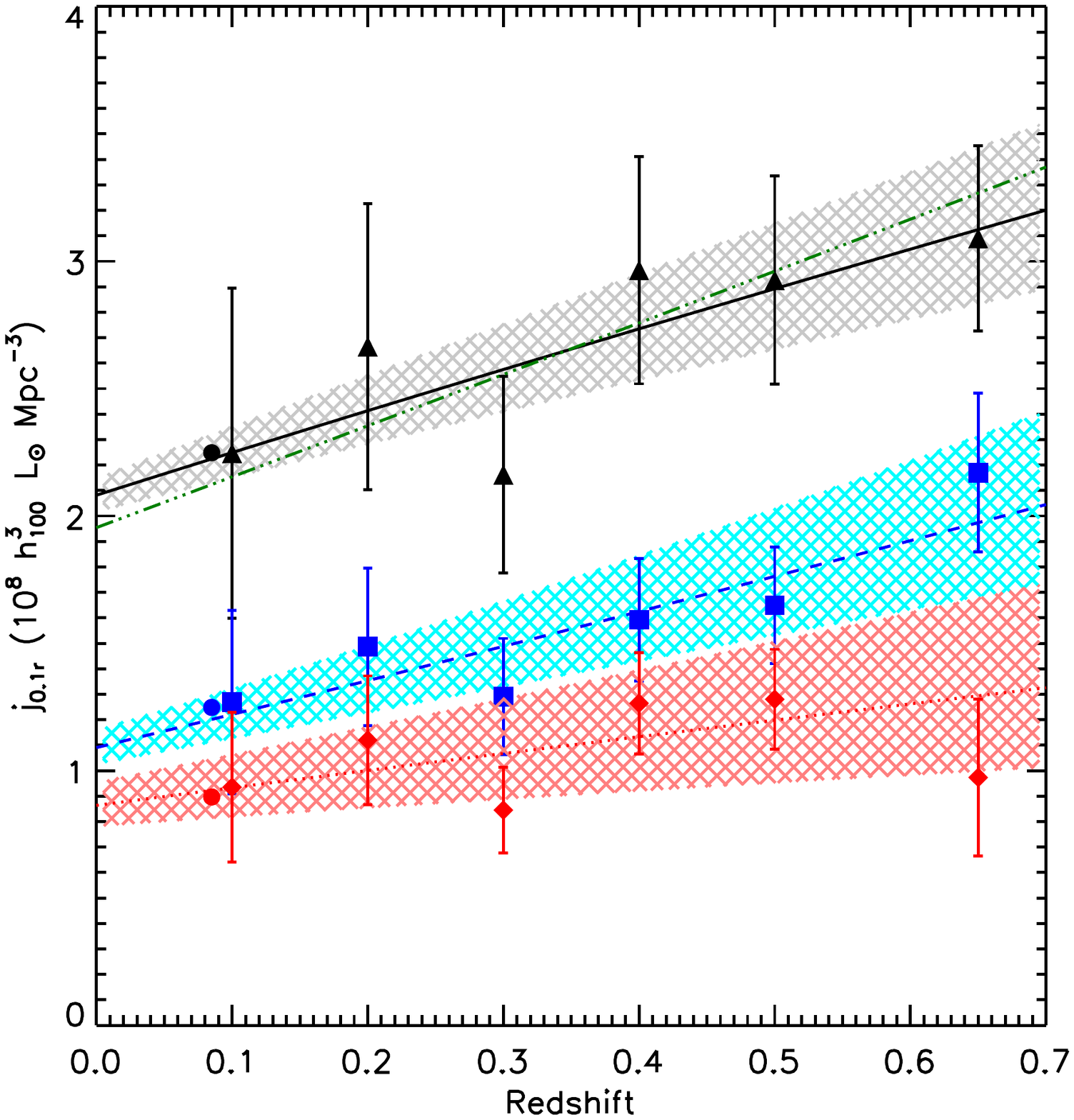}
\caption{\label{fig:rjstar} Evolution of the $^{0.1}r$ luminosity density for all (black triangles), blue (blue squares), and red (red diamonds) galaxies
from AGES.  The SDSS measurement at $z=0.1$ is shown by the filled circles which have been slightly offset in redshift for clarity.  We model the evolution  
as $j\propto (1+z)^n$; the best fitting models are also plotted as the solid, dotted, and dashed lines for the
full, red, and blue samples respectively.  We find that the full sample evolves
with $n=0.81\pm0.27$, the blue galaxies with $n=1.64\pm0.39$, and red galaxies with $n=0.54\pm0.64$. While the number density
of red galaxies has increased since $z=0.7$, the luminosity density in red galaxies grows much more slowly over that time. For comparison, 
the green dot-dot-dot-dashed line shows the sum of the best fitting trends for the red and blue galaxies; the sum of each population 
separately agrees well with the total population within experimental errors.}
\end{figure}

In detail, the variance of the density field of the full region 
is not independent of the variance reported by the jackknife method. This is unavoidable; 
while modes much larger than the survey are invisible to the jackknife method, 
a single large cluster would affect the jackknife errors as well as the overall 
density variation. We remove this double counting by using the Monto Carlo
integration to estimate the covariance matrix between the 15 sub-regions
and predicting the jackknife variations in linear theory.  

If we have $N$ regions on the sky, with areas $A_j$ and a total area of 
$A$, then the covariance between the overdensities, $C_{ij}$of two of the regions is 
\begin{equation}
C_{ij} = 
\int_{\vec\theta_i\in A_i}{d\vec\theta_i\over A_i}
\int_{\vec\theta_j\in A_j}{d\vec\theta_j\over A_j}
w(|\vec\theta_i-\vec\theta_j|)
\end{equation}
where $w(\theta)$ is the angular correlation function.
Constructing a Monte Carlo set of random points in the union of these regions, 
with $N_i$ points in region $i$, the integral can be computed as
\begin{equation}
C_{ij} = {1\over N_i N_j}
\sum_{pq} w(|\vec\theta_p-\vec\theta_q|)
\end{equation}
where the sum is over points $p$ in region $i$ and points $q$ in 
region $j$.  The covariance of the overdensity of the full region is
\begin{equation} 
\sigma_{total}^2 = \sum_{ij} A_i C_{ij} A_j/A^2
\end{equation}
which is the same as if one had integrated the angular correlation function
of pairs of points in the full region.   The variance from jackknife 
is approximately $\sum_{j} C_{jj}/N^2$; this assumes that the regions
are equal in area so as to simplify the formula considerably (i.e.
it assumes that the densities from the $N$ regions are being combined 
with equal weight rather than weighting by area).   This assumption is true
to within 12\% rms for the 15 AGES regions used in our calculations.

Typically, the 
jackknife variance is about 25\% of the total variance.  We then subtract 
this jackknife variance from the full density-field variance to yield the
variance $\sigma_{LSS}^2$ resulting from structure larger than the survey volume. 
Our estimate of the fractional variance in the luminosity density will be the sum
of this variance and the variance from the actual jackknife estimation. By 
subtracting off the linear theory jackknife prediction and adding back the
actual jackknife error, we include the non-linear aspects
of the correlation function and the shot noise to our error estimate. As
the AGES galaxies are not strongly biased relative to the dark matter \citep{Hickox2009}, 
we expect this estimate to be a robust measurement of the variance from large scale
structure.

For our six redshift slices, we find $\sigma_{LSS}/j$ of 0.277, 0.203, 
0.168, 0.144, 0.1128, and 0.079 (from low to high redshift) assuming a linear
theory normalization of $\sigma_8=1.0$ for galaxies.   In all but the 
highest redshift bin, this error is larger  
than the 5-10\% jackknife variance estimate in the luminosity density.  Hence, 
the uncertainty in the estimate of the luminosity density in AGES is dominated by 
large-scale structure.

Figures \ref{fig:rmstar}-\ref{fig:rjstar} show
the redshift dependence of $M_*$, $\phi_*$, and the luminosity density $j$ 
for each of the three galaxy samples in the $^{0.1}r$ band.  Evolution is 
clearly detected.   In all three samples, $M_*$ is brighter in the past.
The slope of this evolution is roughly 1.6 in all three cases. 
 We also find a drop in $\phi_*$ toward higher redshifts,
predominately for the red galaxies. 
For the red galaxies, the rise in $M_*$ balances the drop in $\phi_*$ so
the luminosity density of red galaxies is roughly constant. For
blue galaxies, the
rising $M_*$ dominates and the luminosity density
 increases by 60\% from $z=0.1$ to $z=0.6$.  The luminosity 
density of all galaxies increases by 50\% over the same epoch.   

 Table 
\ref{tab:lf_rparam} lists the best fitting evolution of $M_*$, $\phi_*$, and
$j_{^{0.1}r}$ for the AGES sample.  We assume functional forms such that 
$M_*(z)=M_*(0)+Qz$, $\phi_*\propto 1+Pz$, and $j_{^{0.1}r}\propto(1+z)^n$. 
The table lists the best fitting values for $Q$, $P$, and $n$ for each 
galaxy population. In each of these fits, we perform a 
$\chi^2$ fit of each form to the measured parameters
and associated errors and solve for the best fitting $Q$, $P$, and $n$
value.  While we opt to fit the parameters derived in each redshift bin,
more sophisticated techniques exist allow one to fit the full galaxy
population across all redshifts while simultaneously fitting for the 
evolutationary parameters \citep{Lin99, Heyl1997}.

\subsection{Comparison with Previous Work}
\label{sec:compare}

\begin{figure}
\plotone{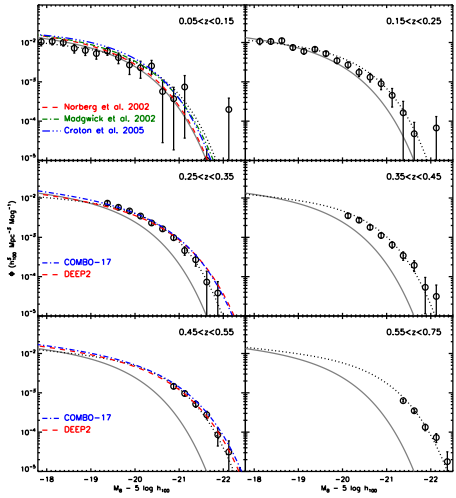}
\caption{\label{fig:ages_blf} AGES $B$-band total galaxy luminosity function.  In each
panel, the data points and errorbars show the $1/V_{max}$ derived luminosity functions, 
the grey line shows the SDSS total galaxy luminosity function and the dotted line shows
the sum of the AGES blue and red galaxy luminosity function.  For comparison, we have
also plotted best fitting luminosity functions from \citet{Norberg02}, \citet{Madgwick2002},
and \citet{Croton2005} from 2dF at low redshift and from DEEP2 and COMBO-17.}
\end{figure}

\begin{figure}
\plotone{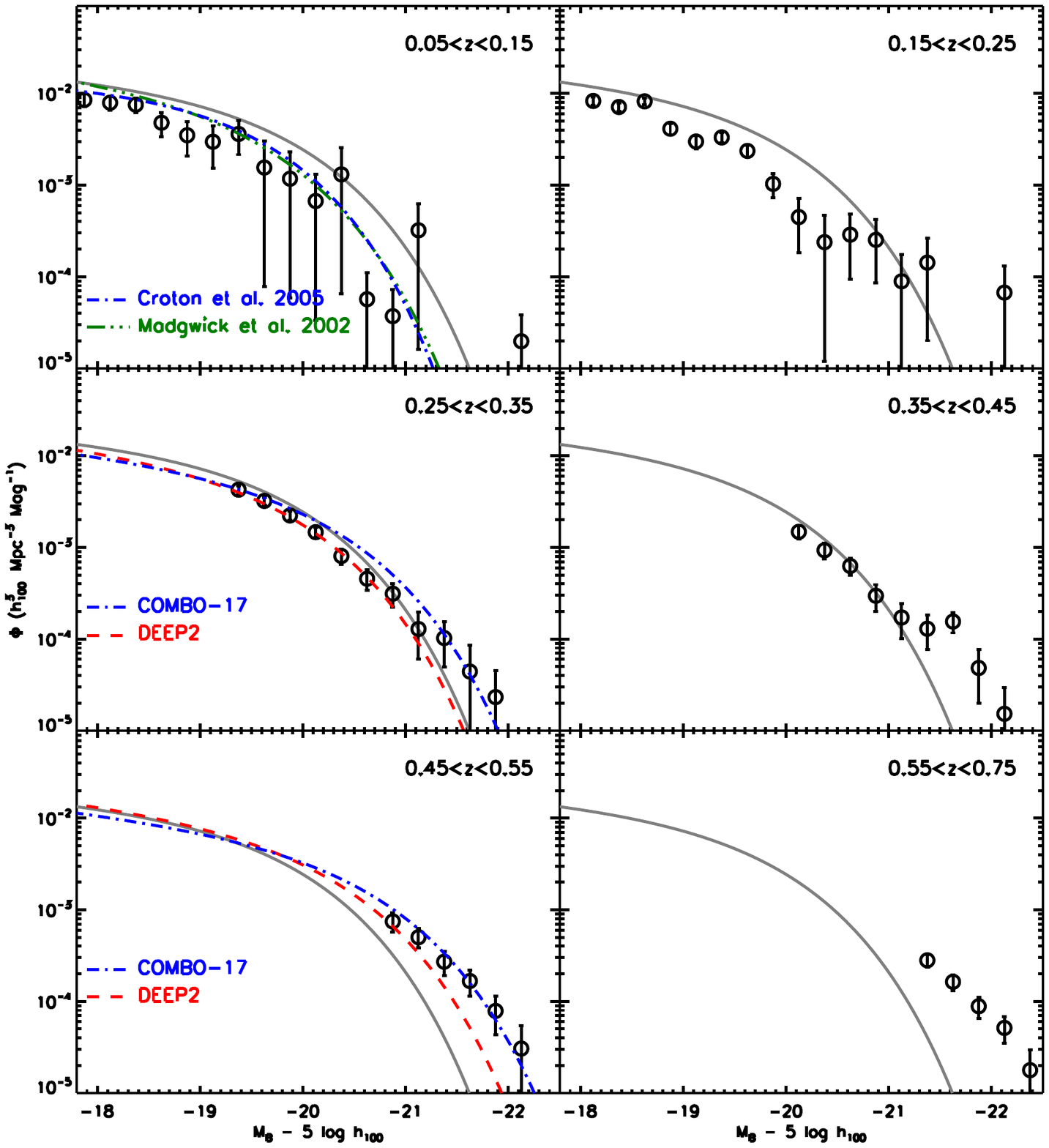}
\caption{\label{fig:ages_blf_blue} Same as \ref{fig:ages_blf} but showing the blue galaxy $B$-band
luminosity function from AGES.}
\end{figure}

\begin{figure}
\plotone{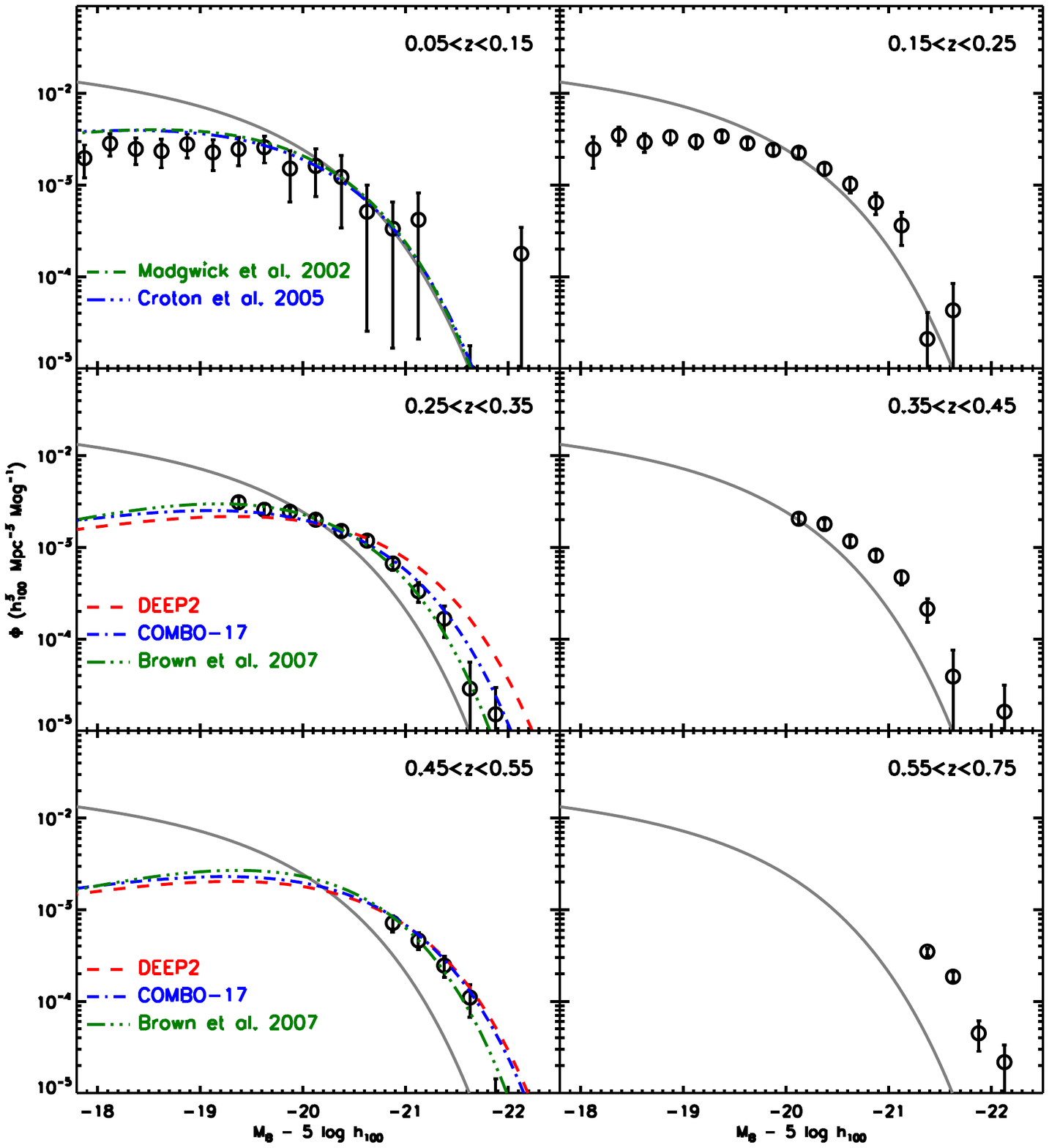}
\caption{\label{fig:ages_blf_red} Same as \ref{fig:ages_blf} but showing the red galaxy $B$-band
luminosity function from AGES.}
\end{figure}

As the majority of work on the galaxy optical luminosity function has focused 
on the restframe $B$-band, we also measure the evolution of the AGES
LF in that band. When constructing the $B$-band luminosity function, 
the observed $R$-band is a close match to the effective wavelength of the rest-frame
$B$-band across the redshift range probed by AGES
(with the closest comparison at $z\sim0.5$).  In order to properly construct the
likelihood when constructing this sample, we implement an effective $R$-band magnitude
cut on the sample to ensure that the $I$-band AGES selection does not bias our sample toward red galaxies.
This effective cut, however, has little impact on our sample and results in the removal  $<0.5\%$ of our
sample galaxies when constructing the $B$-band luminosity function.
We use the SDSS values for $\alpha$ at $z=0.1$ derived from 
the luminosity functions in Figure \ref{fig:sdssBlf} and 
assume the AGES $B$-band luminosity functions have fixed faint-end slopes 
relative to the SDSS value. Figures \ref{fig:ages_blf}-\ref{fig:ages_blf_red} show the AGES
$B$-band luminosity functions compared with luminosity functions from the literature for all galaxies, 
blue galaxies, and red galaxies respectively.  The $1/V_{\mbox{max}}$ luminosity function measurements are presented
in Tables \ref{tab:ages_B_All_lf}-\ref{tab:ages_B_Blue_lf}.
The best fitting STY parameters for the $B$-band luminosity functions are listed in Table \ref{tab:lf_bfits}.

\begin{figure}
\plotone{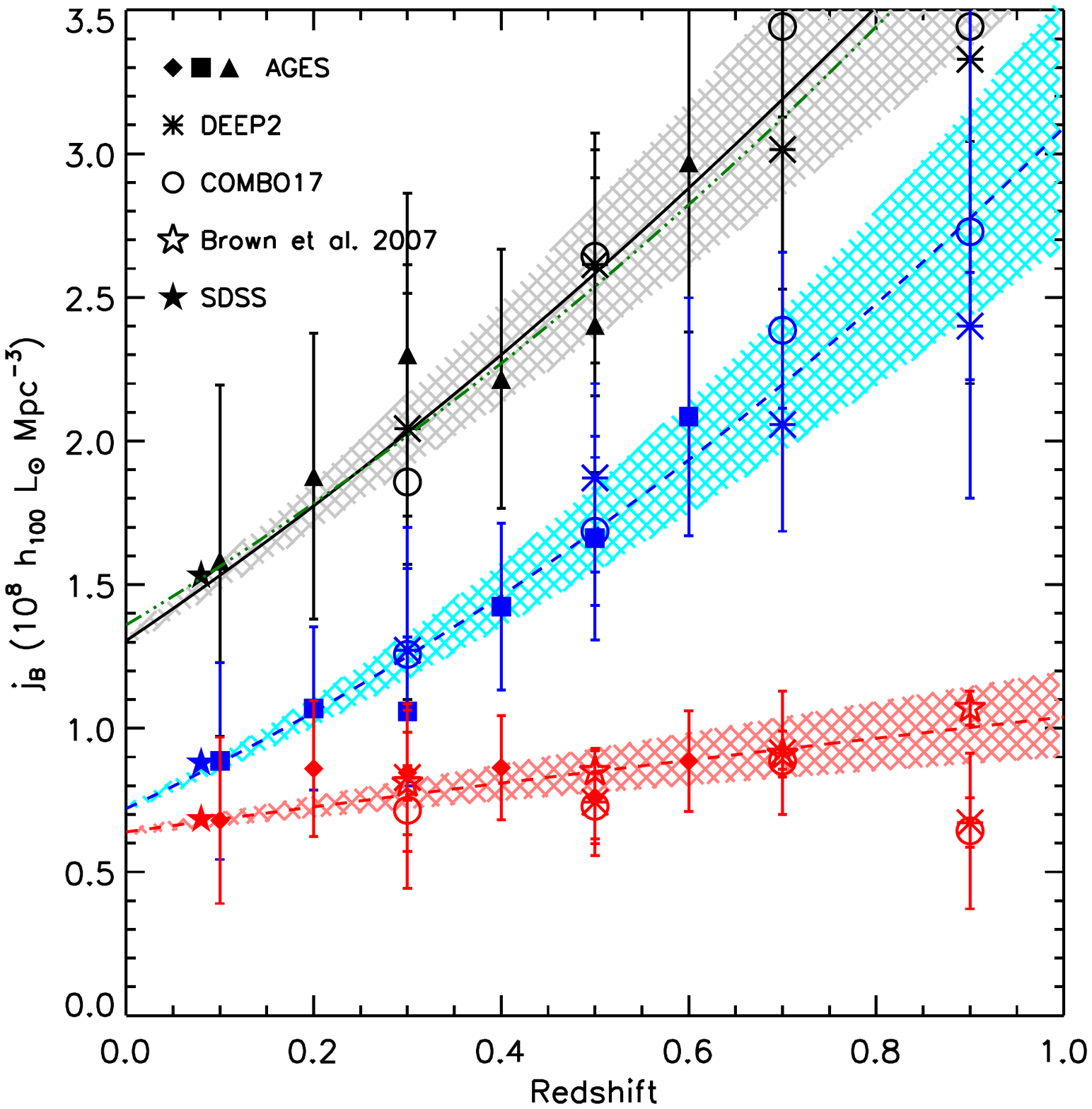}
\caption{\label{fig:ages_B_jden}  The evolution of the $B$-band luminosity density from AGES and the literature. 
 The AGES luminosity density measurements for all galaxies
(solid black triangles), blue galaxies (solid blue squares), and red galaxies (solid red diamonds) is fit using the 
$z=0.1$ SDSS measurements (filled circles) as a required constraint.
We also show the measurements from DEEP2 (asterisks) and COMBO17 (empty circles) for each of the galaxy populations.  
Finally, we also show the estimates of \citet{Brown07} for red galaxies
with stars.   For illustration, we extrapolate our fits at low redshift to $z=1$ and show the $1-\sigma$ confidence as the 
grey, cyan, and pink shaped regions. Overall, the extrapolations of our fits at low-redshift agree well with the measurements 
in the literature at higher redshift.  While current surveys either probe the low-redshift
or high-redshift end of this distribution, with little existing data able to span the entire range and allow for detailed 
measurements within a single survey, overall
the results from high- and low- redshift seem to agree. 
As in Figure \ref{fig:rjstar}, the green dot-dot-dot-dashed line marks
the sum of the best-fitting red and blue luminosity densities.}
\end{figure}

Figure \ref{fig:ages_B_jden} shows the AGES measured B-band 
luminosity density, $j_B$, for all, red, and
blue galaxies as well as measurements for each sample from DEEP2 \citep{Faber07}
and COMBO17 \citep{Wolf03}.  We also include the luminosity density measurements of
\citet{Brown07} which include red galaxies with photometric redshifts also
drawn from the NDWFS field. As the \citet{Brown07} sample covers the same
area based on the same photometric data, it shares the same photometric systematics
and large scale structure systematics as our AGES data.  The dot-dashed lines
and shaded areas show the best fit evolutionary tracks 
for each type of galaxy fit using only the SDSS and AGES data.  
The shaded regions show the $1\sigma$ confidence regions based on the AGES and 
SDSS extrapolated to $z=1$.

Overall, the extrapolations of the AGES evolution agrees  with the 
measurements made in the literature and often lie between the measurements 
from DEEP2 and COMBO-17.  It it worth noting that if our extrapolation is 
extended to $z=1$, the DEEP2 red galaxy luminosity density is quite low 
while the \citet{Brown07} measurement is slightly above our prediction.  Clearly
the ideal manner of studying the full evolution from $z=1$ to today, however, 
is a large deep survey of spectroscopically observed galaxies with the area
and depth to probe the full $z<1$ epoch with robust statistics.

\section{Conclusions}
\label{sec:conclusions}

We have computed the optical luminosity function from the AGN and Galaxy Evolution
Survey sample.  This is the largest spectroscopic sample currently available
of field galaxies at $0.3<z<0.7$. At low redshifts, the luminosity function
from AGES is in excellent agreement with the much larger SDSS dataset.  At 
higher redshift, we see clear evidence for evolution of the luminosity function,
with $M_*$ being brighter at higher redshift. We compute the evolution of the 
luminosity density for the full sample as well as for the populations of  
blue and red galaxies separately.  We find that the evolution of the luminosity density
of red galaxies over the $0.1<z<0.65$ is nearly
 constant in the $^{0.1}r$ band, $(1+z)^{0.54\pm0.64}$, while that of blue galaxies
evolves rapidly, $(1+z)^{1.64\pm0.39}$.  Both blue and red galaxies have
a similar evolution in $M_*$, 1.6 magnitudes per unit redshift at fixed 
$\alpha$. The amplitude of the luminosity function, $\phi_*$, decreases 
with redshift in all cases, but more so for red galaxies.  

The major caveat in these results, aside form the ever-present desire to probe
more survey volume, is that our higher redshift samples include only
fairly luminous galaxies.  Given the observed evolution, the AGES
flux limit reaches $L_*$ at about $z\sim0.53$. Hence, our fits are driven by 
galaxies near or above $L_*$. Any inferences about luminosity densities depend
on the extrapolation to lower luminosity galaxies via a constant $\alpha$.
More data from the next generation of deep spectroscopic surveys which probe
larger volumes and constrain fainter galaxies will allow even tighter constraints
of the evolution of galaxies over the last half of cosmic history.  

\acknowledgments

This work was completed through funding from a National Science
Foundation Graduate Research Fellowship.  Support for this work was provided by NASA through Hubble Fellowship grant HF-01217 awarded by the Space Telescope Science Institute, which is operated by the Associated of Universities for Research in Astronomy, Inc., for NASA, under contract NAS 5-26555.  S.Murray, C. Jones, W. Forman, and R. Hickox acknowledge support from NASA Contract NAS 8-03060 and the Chandra GTO Program.

Observations reported here were obtained at the MMT Observatory,
a joint facility of the Smithsonian Institution and the University
of Arizona.  This work made use of images and/or data products
provided by the NOAO Deep Wide-Field Survey \citep{jannuzidey1999},
which is supported by the National Optical Astronomy Observatory
(NOAO). NOAO is operated by AURA, Inc., under a cooperative agreement
with the National Science Foundation.  This work is based in part on observations 
made with the {\it Spitzer Space Telescope,} which is operated by the Jet Propulsion 
Laboratory, California Institute of Technology under a contract with NASA. Support 
for this work was provided by NASA through an award issued by JPL/Caltech.

Funding for the SDSS and SDSS-II has been provided by the Alfred P. Sloan Foundation, the Participating Institutions, the National Science Foundation, the U.S. Department of Energy, the National Aeronautics and Space Administration, the Japanese Monbukagakusho, the Max Planck Society, and the Higher Education Funding Council for England. The SDSS Web Site is http://www.sdss.org/.

The SDSS is managed by the Astrophysical Research Consortium for the Participating Institutions. The Participating Institutions are the American Museum of Natural History, Astrophysical Institute Potsdam, University of Basel, University of Cambridge, Case Western Reserve University, University of Chicago, Drexel University, Fermilab, the Institute for Advanced Study, the Japan Participation Group, Johns Hopkins University, the Joint Institute for Nuclear Astrophysics, the Kavli Institute for Particle Astrophysics and Cosmology, the Korean Scientist Group, the Chinese Academy of Sciences (LAMOST), Los Alamos National Laboratory, the Max-Planck-Institute for Astronomy (MPIA), the Max-Planck-Institute for Astrophysics (MPA), New Mexico State University, Ohio State University, University of Pittsburgh, University of Portsmouth, Princeton University, the United States Naval Observatory, and the University of Washington.

\appendix
\section{The AGES Selection Completeness Function}
\label{sec:galsec}
\subsection{AGES Galaxy Target Selection}

In this section, we outline the galaxy target selection process employed by AGES.  AGES
also targeted quasars, but we will not describe that selection, here; full details on the
AGES execution and target selection can be found in \citet{Kochanek2011} and \citet{Assef2010}.

We targeted based on the NDWFS DR3 imaging catalogs.
For each band of NDWFS imaging, we define acceptable photometry (\bgood, \rgood, and \igood) 
the Sextractor flag $\texttt{FLAG}<8$ (i.e. unsaturated, not falling off a chip boundary, or heavily blended), 
not flagged as a duplicate object, and which had photometric data available ($\texttt{FLAG\_PHOT}=0$). All targets
were required to have \igood\, and at least one of \rgood\ or \bgood.  Objects were classified as point
sources if the stellarity index (Sextractor's $\texttt{CLASS\_STAR}$ parameter)
in at least one of the bands had $\texttt{CLASS\_STAR}\ge0.8$. Galaxy 
targets are required to be extended by this criterion.  All flags discussed here use the default
definition provided by Sextractor.

Galaxy targets were restricted to lie in the $15.45 < I \le 20.45$ magnitude range.  
To avoid the problem
of Kron-like ($\texttt{AUTO}$) magnitudes being corrupted by the halos of nearby stars, we impose 
limits on the aperture magnitudes of selected objects.  If either of the $1\farcs0$ or $6\farcs0$ aperture 
magnitudes were extremely faint,  $I_{1\farcs0}>24.45$ or $I_{6\farcs0}>21.45$, 
we removed the galaxy from the sample. These restrictions remove much of the low-surface brightness 
spurious objects, but not completely.  
In order to correct for these low-surface brightness contaminants, we first flag objects which are
observed near bright USNO stars.  For each USNO star in the NDWFS field, we define a scale length
\begin{equation}
\label{equ:USNObright}
\theta_{\rm USNO} = 20.0'' + 5.0''(15.45-R_{\rm USNO}).
\end{equation}
If the closest USNO star is less than $\theta_{\rm USNO}$ from a galaxy then the galaxy was flagged
as being too close to the bright star.  
If  the $6''$ galaxy aperture magnitude satisfied 
\begin{equation}
\label{equ:USNOquad}
I_{6''} > I_{tot} + 4[(I_{tot}-20.45)/8]^2
\end{equation}
the galaxy was rejected. Secondly,
if the galaxy was less than $0.5\theta_{\rm USNO}$ from the closest USNO star, the galaxy was removed from 
the survey.   After these cuts, we have a cleaned sample of possible galaxy targets.
Table \ref{tab:selection} lists the bright sample magnitude range, faint sample magnitude range, 
and faint sample sampling rate for each of the samples defined for AGES spectroscopy and Table \ref{tab:samplesize} 
lists the number of targets, number of redshifts obtained, and overall completeness for each AGES galaxy sample.
It's 
important to reiterate that all of these are cuts {\it in addition} to the $I<20.45$ cut; 
in essence, a galaxy that was a bright detection from our multiwavelength imaging was given
higher preference for spectroscopy than galaxies undetected in these other bands.

\subsection{Completeness Corrections}

\label{sec:completenesscorrections}
We decompose the selection function into 4 terms.  First, objects may not have 
passed target selection cuts due to quirks of the photometry
or some aspect of the targeting (``photometric incompleteness'').  Second, 
objects that would otherwise have been selected may have been dropped from our
statistical sample due to {\it a priori} sparse sampling.  Third, a high-priority
object may have failed to have a fiber allocated (``fiber incompleteness'').  Finally
a spectrum may have failed to yield a useful redshift (``redshift incompleteness'').  

The AGES galaxy target selection sets flux limits in 12 bands of photometry.
  However, this complex set of targeting criteria can be 
thought of as a simple $I<20.45$ sample in which {\it a priori} sparse sampling
has been done to de-emphasize ``common'' objects while more heavily sampling the 
tails of the multicolor distribution.    It is easy to undo the sparse sampling
and restore a fair $I<20.45$ sample as all of the targeting weights are known 
exactly.  We do not use objects that were rejected by sparse sampling in our analysis
though some of these objects did get a spectrum as a ``filler'' object. 
We construct the main statistical galaxy sample as the objects with high priority target
flags, with all three optical bands present (\rgood, 
\igood, and \bgood) to ensure good $k$-corrections, and inside the primary galaxy survey region.

About 1.5\% of objects fail either \rgood\, or \bgood.  There is little coherent structure to 
these, so it is not a problem of non-overlapping photometry. However, these objects also have
a 25\% redshift failure rate, which is quite high compared to the full sample.  It is 
likely these objects are corrupt; we estimate that the requirements of 3 bands
of good photometry is likely only causing a 1\% photometric incompleteness.  Furthermore, we
rejected galaxies from AGES if they were too close to a bright star; this affects about 
2\% of the survey region.
There are 762 SDSS MAIN sample targets ($r<17.77$)
inside the primary survey region.  733 (96\%) of these are in the AGES parent sample and 709 (93\%)
end up being selected.  The 3\% loss is due to the effects described above.  However, the 
first 4\% is yet unaccounted for.  Of course, the SDSS has a small rate of spurious targets, about
1\%. We therefore conclude that at $r<17.77$, we have an additional 3\% incompleteness that 
we have not identified.  This could well improve as one moves to the fainter objects that NDWFS 
was designed for.  In order to ensure that this 3\% incompleteness is appropriate for the full 
galaxy sample and not simply the bright end of the galaxy luminosity function which overlaps with the
SDSS MAIN sample, we constructed fake galaxies with sizes and fluxes representative of the AGES
galaxy sample and included them in the NDWFS imaging.  Reperforming a SEXTRACTOR analysis of the imaging, 
we find that approximately 3.5\% of the galaxies added to the images are unrecovered - typically due to deblending
with bright foreground galaxies or stars.  This incompleteness value does not strongly depend on the brightness 
of the galaxy within the range of galaxy fluxes considered for our luminosity function analysis. 
\citet{Brown07} found the NDWFS imaging dataset to be more than 85\% complete for galaxies of 
typical sizes and shapes to $I_{AB}=23.7$, several magnitudes deeper than we target with AGES.  
As the optical imaging used to select AGES target galaxies extends several magnitudes deeper than the
flux limit of AGES spectroscopy, we expect little incompleteness to arise from photometric depth at the faint 
end of our targeting range.

In sum, we believe that the photometric incompleteness is 3-6\% (but at the very bright 
end, e.g. $I<15$, there is surely more incompleteness).  The primary galaxy survey region
is 7.90 deg$^2$ We adopt a 4\% catalog incompleteness and adjust the fiducial area to 7.60 deg$^2$.
The large-scale structure corrections remain tied to the larger area because the incompleteness
is in many small disjoint regions.  

In order to test the effects of star-galaxy separation on our target selection, we first
compare the stellarity classifications between NDWFS and SDSS imaging.  We restrict this comparison
to bright SDSS objects with robust stellarity measurements; as the depth of NDWFS is considerably 
deeper than SDSS, extending to the detection limit of SDSS leads to identifying problems with the 
SDSS star-galaxy separation at faint fluxes rather than to understand the role of NDWFS star-galaxy
separation on our luminosity function measurements.  The NDWFS photometry reproduce the stellarity 
measurements from SDSS in this regime.  In order to quantitatively test how the separation of stars 
galaxies effects fainter NDWFS targets, we select all objects with IRAC colors consistent with 
arising from galaxies (and inconsistent with being either stars or AGNs).  We find that $<0.3\%$
of these objects were classified as stars based on their NDWFS stellarity measurements and conclude
that our samples do not suffer significant incompleteness due to galaxies becoming poorly resolved
in the NDWFS imaging.

Only 1\% of the AGES main-sample galaxies lack counterparts in SDSS imaging.  Nearly all of these 
are low surface brightness objects that are plausibly below the SDSS detection limit.  Thus, we
can limit the rate of spurious objects in the AGES sample to below 1\%.  In 2005, AGES targeted
nearly 300 bright objects, typically $r<17$, that SDSS declared as stars (which is confirmed by 
AGES spectroscopy). These objects saturated the NDWFS imaging and were mistakenly classified as 
extended by the star-galaxy separation.  To remedy this, we remove all objects at
$r<19$ that SDSS called a point source.  NDWFS and SDSS agree very well on extended sources 
between $r=17$ and $r=20$.  
SDSS obtained redshifts for 27 galaxies included in the AGES primary sample that we didn't observe
at the MMT.  We consider these galaxies as having good redshifts when computing fiber and redshift
incompleteness.

\begin{figure}
\includegraphics[width=5in, height=5in]{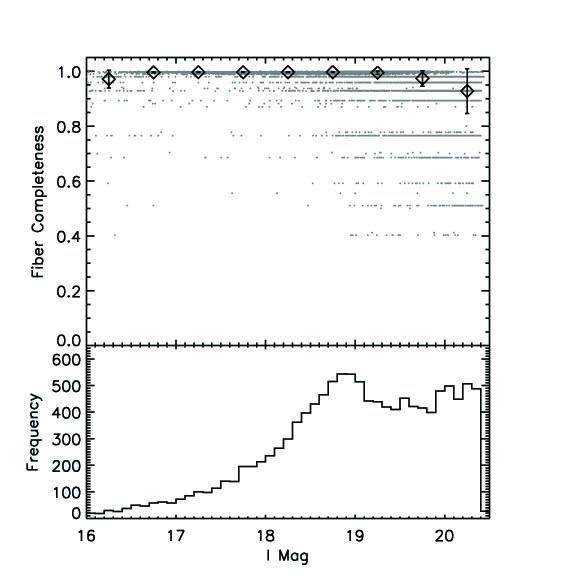}
\caption{\label{fig:fibercompleteness} Final AGES fiber completeness
as a function of total $I$-band magnitude.  This value represents the correction 
applied to each galaxy due to observed objects which did not receive a fiber
due to fiber collisions as described in \S\ref{sec:completenesscorrections}.}
\end{figure}

\begin{figure}
\plotone{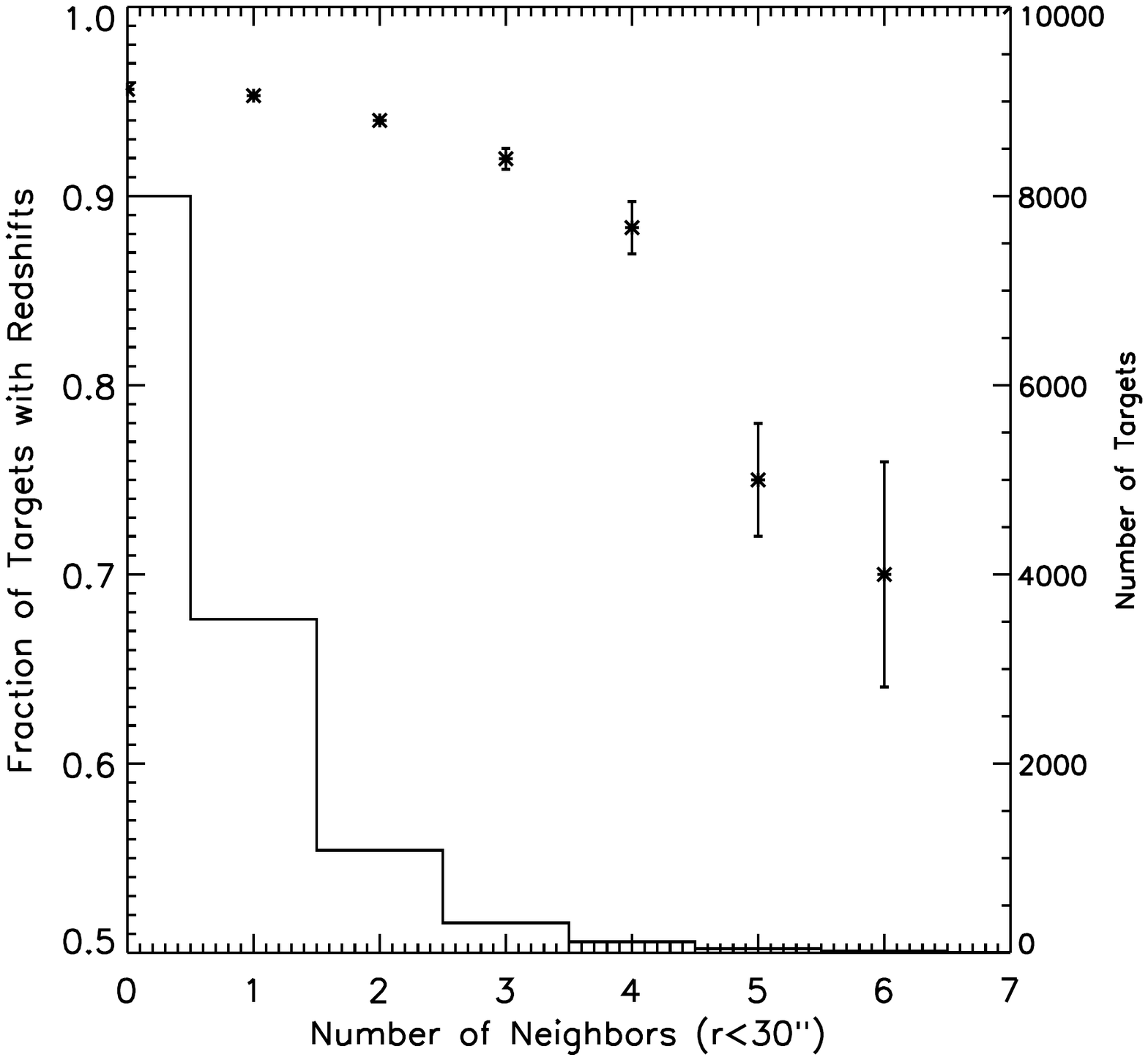}
\caption{\label{fig:ages_neighborcomp} Fraction of AGES targets which received redshifts as a function 
of the number of neighbor galaxies within 30'' of the target's location. The histogram (right axis)
shows the distribution of neighbor counts for all of the AGES targets. When calculating the completeness 
correction for each galaxy in our survey we include the effects of neighboring galaxies as described in the text.}
\end{figure}

AGES performs very well as regards to fiber completeness; most regions of the survey
were observed at least six times, so even high multiplet groupings could be resolved.
In total, 95.7\% of targets were observed.  Fields 13-15, however, were not as well sampled in 
2005. Field 14 is the worst with only 83\% completeness.  Fields 1-12 have fiber 
completeness of 98.3\%.  
The level of fiber incompleteness depends on the local environment of
the galaxy, as fiber collisions are a major source of the problem (though not the only source).
To address this, we seek to assess completeness as a function of local density and 
number of opportunities for each object to be observed. We first split the objects
into those that were high-priority targets in 2004 and those that were not.  The 2004 
targets have a much higher completeness and are thus treated separately from objects
observed in 2005 and 2006.  For each object, we count the number of high-priority targets within
$30''$.  For the 2004 targets, we divide them into sets according to the neighbor count and
define the fiber completeness as the fraction of objects in each bin that received a fiber.
For the 2005 targets, we repeat this but restrict the count to high-priority targets that
ere not observed in 2004. We also separate the binning into objects that fell within the 
field of $<4$, 4, 5, or $>5$ observations in 2005 and 2006.   In other words, the fiber completeness
of 2005 targets is judged in sets defined by the number of 2005 nights and the number of 2005
and 2006 opportunities.  In the extremes of the distribution, a bin can have zero observed galaxies
and yet not be empty from non-observed galaxies, which would result in an infinite weight.  In these
cases, we add objects to the next bin until we have a galaxy that can be up-weighted to 
compensate for the missing ones.  Figure \ref{fig:fibercompleteness} shows the distribution of
the final fiber completeness correction derived here as a function of target total magnitude and 
Figure \ref{fig:ages_neighborcomp} shows the mean fraction of possible AGES targets that received a valid redshift
as a function of the number of nearby neighbor galaxies.

We would have preferred to define the fiber completeness more locally, e.g., count neighbors and
find the fraction that got observed, as this would tie the incompleteness closer to the large-scale
structure.  This does not work because some rather isolated targets failed to get fibers, 
generally due to high over density elsewhere in the field, and so we are left with unobserved 
galaxies with undefined weights. This is only a problem in the less complete fields (13-15).

\begin{figure}
\plotone{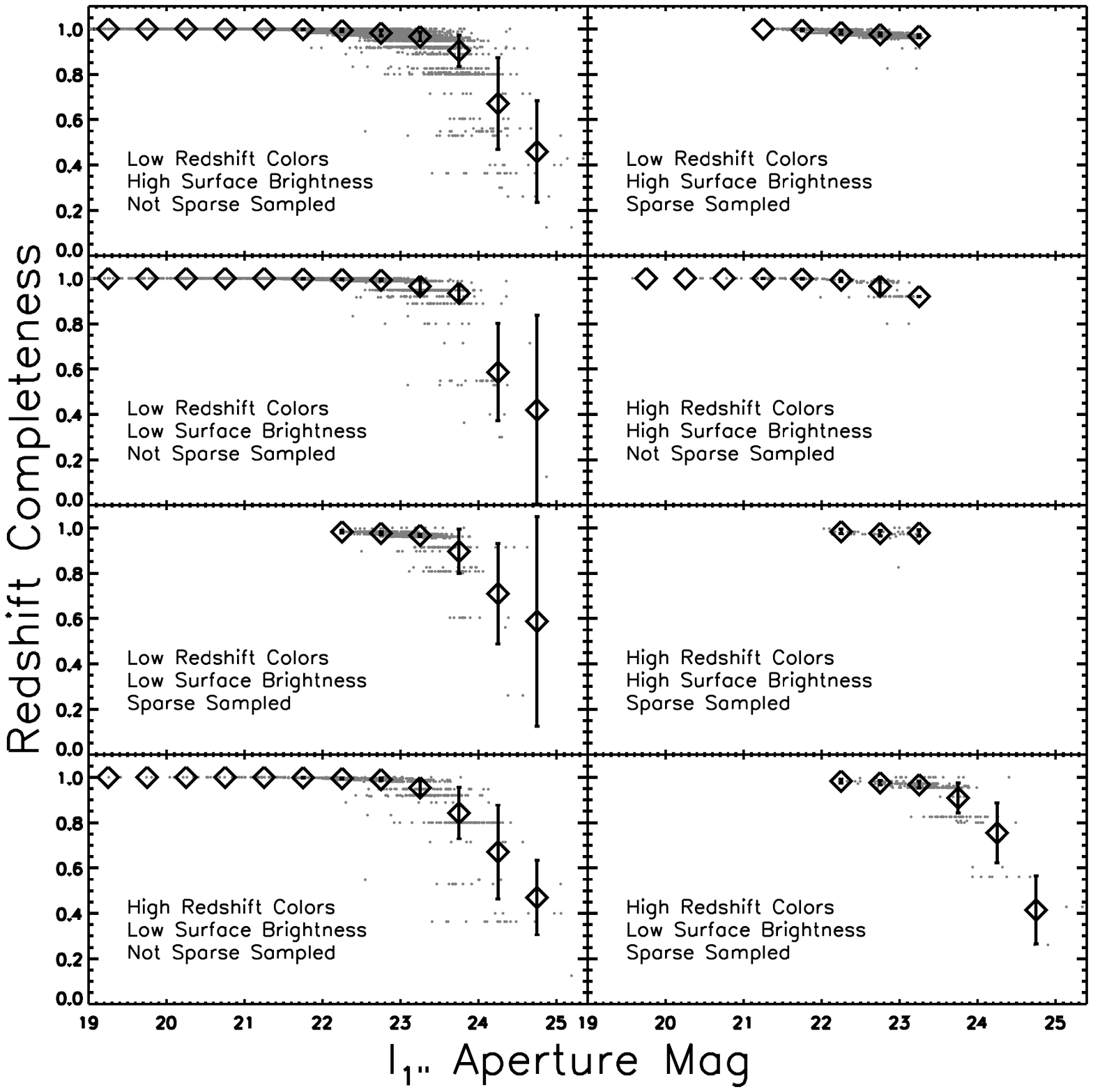}
\caption{\label{fig:redshiftcompleteness} Redshift completeness as a function of 
$I$-band aperture magnitude (measuring the flux contained within a Hectospec fiber).
When calculating this completeness, we consider objects based on three criteria with
limits described in the text.  First, we separate galaxies based on their observed colors
to separate galaxies with high- and low-redshift colors.  We secondly divide the sample based
on the surface brightness of each galaxy. Finally, we calculate the completeness separately for 
galaxies which were sparse sampled and those that were not.  This separation gives rise to 
each of the 8 panels in the Figure.  The subdivision of each sample is listed in the
lower-left and the mean trend is showed by the diamonds with errorbars illustrating the 
1-$\sigma$ range in values.  When calculating the luminosity function, we utilize the
raw completeness values in the figure; the means are shown to guide the eye.}
\end{figure}

For redshift completeness, we are primarily concerned about trends with surface brightness. There
are 274 primary galaxy targets (2.1\%) that received a fiber but failed to get a redshift (of course, 
there are many more failed observations, but we reobserved most of them in order to get a
useful redshift).  We track the surface brightness by the $I$-band $1''$ aperture magnitude. 
In order to explore the redshift rate success, we apply three criteria to our sample:
\begin{enumerate}
\item Objects faintward of the main surface brightness locus as 
defined by $I_{\rm1\farcs0}-0.8 \Itot>7$, or not.
\item Objects whose colors suggest high redshift, 
$(R-I)-0.2(B_W-R)>0.55$, or not. \
\item Objects that were sparse sampled.
\end{enumerate}
Based on an object passing or failing each of these criteria, we derive 8 sets of objects.
In each set, we bin the galaxies in 0.5 mag bins in aperture 
magnitude and consider the rate of getting a successful redshift (among the targets that received a 
fiber).   Figure \ref{fig:redshiftcompleteness} shows the redshift completeness derived using this
method as a function of the aperture magnitude of each AGES galaxy target.
It is worth noting that we have not explicitly considered the signal-to-noise ratio of the
observations in this spectral completeness model. In practice, some observations
were better than others, and this would alter the angular structure of the corrections, but 
we believe this to be a minor effect. 

\begin{figure}
\plotone{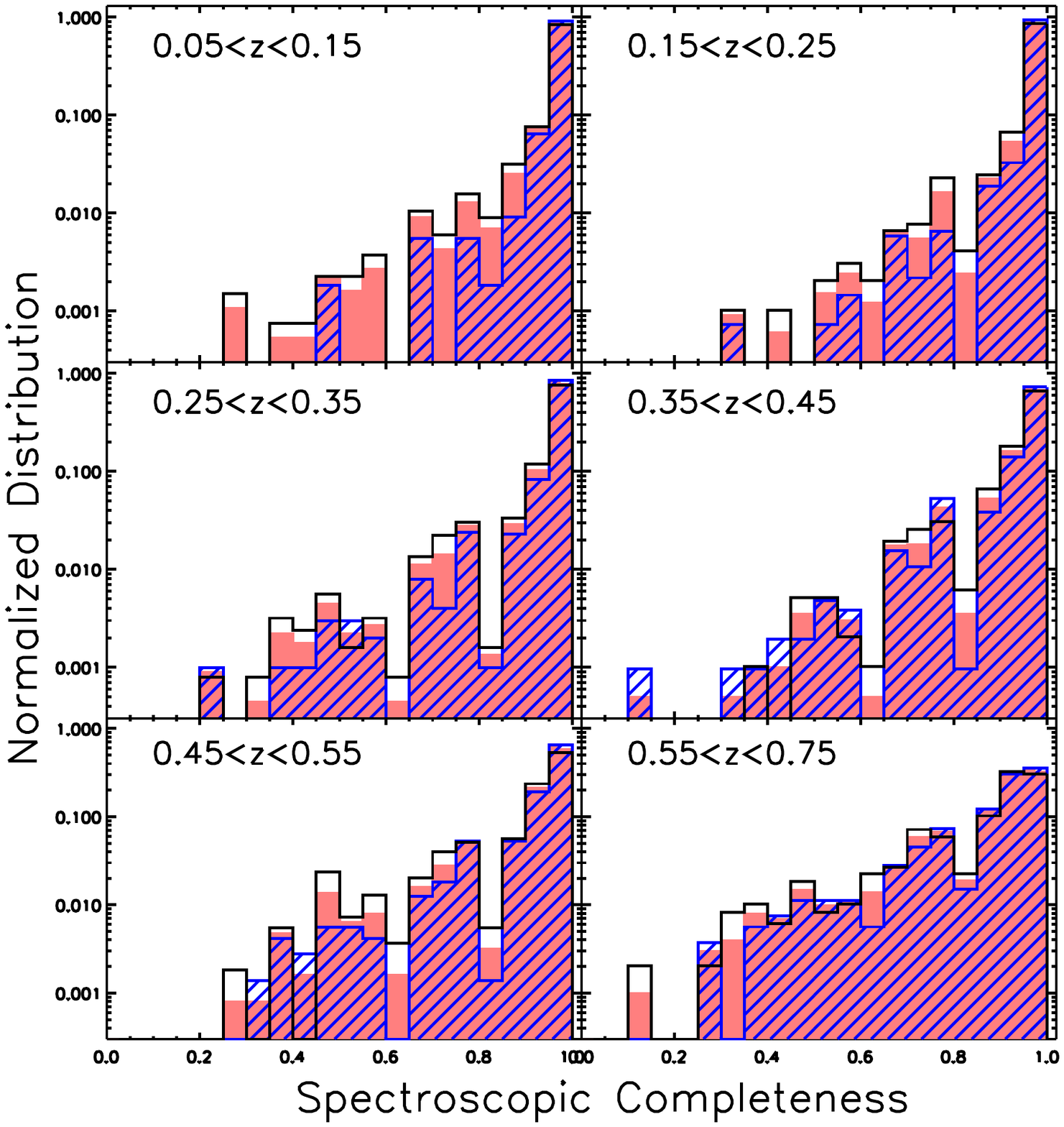}
\caption{\label{fig:compdistro} Distribution of final spectroscopic completeness (not including 
corrections due to sparse sampling) for the full AGES galaxy sample.  The empty histogram
shows the full distribution of the sample.  The filled histogram shows the distribution for 
red galaxies while the hashed histogram shows that for blue galaxies.}
\end{figure}

Our final galaxy weight is calculating by multiplying the inverse of the sparse sampling rate, the
fiber completeness, and the redshift completeness.   Figure \ref{fig:compdistro} illustrates 
the distribution of spectroscopic weight (the product of the fiber weight and redshift completeness weight)
for the full galaxy sample as well as the red and blue galaxy samples separately. 
We have corrected for the photometric 
incompleteness by correcting the effective survey area as described above.   In total, 
the main galaxy sample (after sparse sampling, area restrictions, and applying a
$I_{\rm tot} < 20.4$ flux limit) has 12,473 objects with good redshifts. Summing the weights yields
25,972 effective objects. This is in good agreement (0.3\%) with the 26,033 targets in the parent
$I_{\rm tot} < 20.4$ sample after requiring good photometry in $B_W$, $R$, and $I$ and 
excluding bright stars from SDSS. These numbers do not match exactly because the sparse sampling 
was random and because of small unmodeled interactions between the various terms of the 
selection function (e.g., the fiber incompleteness differs slightly from one class of 
sparse-sampling to the next, but we have assumed the two independent when we multiply the corrections). 
The redshift distribution of AGES galaxies is shown in Figure \ref{fig:zhist}.

In summary, AGES successfully observed half of the total $I_{\rm tot}<20.4$ photometric sample in the
\bootes field.  Nearly all of this was due to the {\it a priori} sparse sampling which can be 
corrected exactly.  The fiber incompleteness is 4.3\% on average; the redshift incompleteness is
2.1\%.  Given the very high completeness, we are confident that the first-order attempts described
above to correct the lingering incompleteness reduce the completeness uncertainties to well below the statistical
uncertainties.

\begin{deluxetable}{crrrrrr}
\tablecolumns{7}
\tablewidth{0pt}
\tabletypesize{\scriptsize}
\tablecaption{AGES $^{0.1}$r-band $1/V_{\rm max}$ Luminosity Functions for All Galaxies \label{tab:ages_r_All_lf}}
\tablehead{
\colhead{} & 
\multicolumn{6}{c}{Luminosity Function ($\times 10^{-4}$ $h^3$ Mpc$^{-3}$ mag$^{-1}$)}\\
\colhead{Luminosity Range\tablenotemark{a}} & 
\colhead{$0.05<z<0.15$} & 
\colhead{$0.15<z<0.25$} &
\colhead{$0.25<z<0.35$} &
\colhead{$0.35<z<0.45$} & 
\colhead{$0.45<z<0.55$} & 
\colhead{$0.55<z<0.75$}}
\startdata
$-18.25<M<-18.00$ & $140.82\pm17.07$ & \nodata & \nodata & \nodata & \nodata & \nodata\\
$-18.50<M<-18.25$ & $133.97\pm09.78$ & \nodata & \nodata & \nodata & \nodata & \nodata\\
$-18.75<M<-18.50$ & $113.45\pm08.81$ & $131.17\pm05.82$ & \nodata & \nodata & \nodata & \nodata\\
$-19.00<M<-18.75$ & $106.26\pm09.99$ & $109.51\pm04.64$ & \nodata & \nodata & \nodata & \nodata\\
$-19.25<M<-19.00$ & $119.46\pm10.11$ & $121.39\pm18.13$ & \nodata & \nodata & \nodata & \nodata\\
$-19.50<M<-19.25$ & $106.22\pm08.42$ & $108.94\pm04.67$ & \nodata & \nodata & \nodata & \nodata\\
$-19.75<M<-19.50$ & $ 92.76\pm07.83$ & $108.32\pm04.80$ & \nodata & \nodata & \nodata & \nodata\\
$-20.00<M<-19.75$ & $ 88.19\pm07.64$ & $ 99.08\pm04.45$ & $ 70.31\pm02.73$ & \nodata & \nodata & \nodata\\
$-20.25<M<-20.00$ & $ 90.35\pm10.75$ & $ 98.19\pm04.80$ & $ 58.42\pm02.65$ & \nodata & \nodata & \nodata\\
$-20.50<M<-20.25$ & $ 52.88\pm05.99$ & $ 70.12\pm03.68$ & $ 56.53\pm02.37$ & \nodata & \nodata & \nodata\\
$-20.75<M<-20.50$ & $ 49.64\pm05.90$ & $ 59.62\pm03.39$ & $ 45.67\pm02.28$ & $ 56.81\pm01.86$ & \nodata & \nodata\\
$-21.00<M<-20.75$ & $ 38.72\pm05.29$ & $ 41.66\pm03.59$ & $ 34.54\pm01.90$ & $ 50.26\pm01.76$ & \nodata & \nodata\\
$-21.25<M<-21.00$ & $ 30.29\pm04.74$ & $ 33.31\pm02.53$ & $ 29.82\pm01.69$ & $ 37.48\pm01.52$ & \nodata & \nodata\\
$-21.50<M<-21.25$ & $ 20.45\pm04.90$ & $ 24.80\pm02.18$ & $ 18.55\pm01.34$ & $ 25.57\pm01.26$ & $ 29.04\pm01.12$ & \nodata\\
$-21.75<M<-21.50$ & $ 12.68\pm03.48$ & $ 14.39\pm01.66$ & $ 11.06\pm01.03$ & $ 17.00\pm01.01$ & $ 24.08\pm01.02$ & \nodata\\
$-22.00<M<-21.75$ & $  5.75\pm02.59$ & $  7.14\pm01.17$ & $  7.20\pm00.83$ & $  9.69\pm00.76$ & $ 13.61\pm00.77$ & $ 14.28\pm00.61$\\
$-22.25<M<-22.00$ & $  4.77\pm03.37$ & $  3.49\pm00.82$ & $  3.61\pm00.59$ & $  6.19\pm00.61$ & $  9.22\pm00.63$ & $ 10.12\pm00.49$\\
$-22.50<M<-22.25$ & $ 26.57\pm25.07$ & $  2.45\pm00.69$ & $  1.66\pm00.40$ & $  2.43\pm00.38$ & $  5.23\pm00.48$ & $  8.67\pm00.46$\\
$-22.75<M<-22.50$ & $  0.67\pm00.66$ & $  1.39\pm00.52$ & $  0.77\pm00.27$ & $  2.13\pm00.36$ & $  2.40\pm00.32$ & $  3.36\pm00.28$\\
$-23.00<M<-22.75$ & $  0.69\pm00.67$ & $  0.54\pm00.47$ & $  0.19\pm00.14$ & $  0.18\pm00.10$ & $  1.41\pm00.25$ & $  2.10\pm00.22$
\enddata
\tablenotetext{a}{$M=M_{^{0.1}r}-5 {\rm log} h$}
\end{deluxetable}

\begin{deluxetable}{crrrrrr}

\tablecolumns{7}
\tablewidth{0pt}
\tabletypesize{\scriptsize}
\tablecaption{AGES $^{0.1}$r-band $1/V_{\rm max}$ Luminosity Functions for Blue Galaxies \label{tab:ages_r_Blue_lf}}
\tablehead{
\colhead{} & 
\multicolumn{6}{c}{Luminosity Function ($\times 10^{-4}$ $h^3$ Mpc$^{-3}$ mag$^{-1}$)}\\
\colhead{Luminosity Range\tablenotemark{a}} & 
\colhead{$0.05<z<0.15$} & 
\colhead{$0.15<z<0.25$} &
\colhead{$0.25<z<0.35$} &
\colhead{$0.35<z<0.45$} & 
\colhead{$0.45<z<0.55$} & 
\colhead{$0.55<z<0.75$}}
\startdata
$-18.25<M<-18.00$ & $123.12\pm16.69$ & \nodata & \nodata & \nodata & \nodata & \nodata\\
$-18.50<M<-18.25$ & $113.97\pm09.08$ & \nodata & \nodata & \nodata & \nodata & \nodata\\
$-18.75<M<-18.50$ & $ 84.64\pm07.50$ & $108.12\pm05.36$ & \nodata & \nodata & \nodata & \nodata\\
$-19.00<M<-18.75$ & $ 75.80\pm07.47$ & $ 77.57\pm03.93$ & \nodata & \nodata & \nodata & \nodata\\
$-19.25<M<-19.00$ & $ 87.82\pm07.77$ & $ 96.07\pm17.99$ & \nodata & \nodata & \nodata & \nodata\\
$-19.50<M<-19.25$ & $ 68.98\pm06.81$ & $ 75.58\pm03.92$ & \nodata & \nodata & \nodata & \nodata\\
$-19.75<M<-19.50$ & $ 61.50\pm06.37$ & $ 65.88\pm03.64$ & \nodata & \nodata & \nodata & \nodata\\
$-20.00<M<-19.75$ & $ 51.44\pm05.83$ & $ 58.16\pm03.35$ & $ 47.53\pm02.29$ & \nodata & \nodata & \nodata\\
$-20.25<M<-20.00$ & $ 45.16\pm06.23$ & $ 56.42\pm03.68$ & $ 31.45\pm01.74$ & \nodata & \nodata & \nodata\\
$-20.50<M<-20.25$ & $ 27.14\pm04.27$ & $ 35.96\pm02.63$ & $ 34.90\pm01.83$ & \nodata & \nodata & \nodata\\
$-20.75<M<-20.50$ & $ 22.75\pm03.96$ & $ 28.86\pm02.36$ & $ 24.46\pm01.78$ & $ 30.26\pm01.35$ & \nodata & \nodata\\
$-21.00<M<-20.75$ & $ 17.45\pm03.53$ & $ 20.27\pm02.96$ & $ 16.18\pm01.27$ & $ 22.28\pm01.17$ & \nodata & \nodata\\
$-21.25<M<-21.00$ & $ 10.34\pm02.70$ & $ 15.27\pm01.71$ & $ 13.55\pm01.14$ & $ 18.43\pm01.07$ & \nodata & \nodata\\
$-21.50<M<-21.25$ & $  9.38\pm02.75$ & $ 10.64\pm01.43$ & $  7.92\pm00.87$ & $  9.24\pm00.75$ & $ 13.59\pm00.77$ & \nodata\\
$-21.75<M<-21.50$ & $  3.47\pm01.71$ & $  5.05\pm00.98$ & $  5.14\pm00.70$ & $  5.67\pm00.58$ & $  9.17\pm00.63$ & \nodata\\
$-22.00<M<-21.75$ & $  0.70\pm00.68$ & $  1.17\pm00.48$ & $  2.66\pm00.50$ & $  3.97\pm00.49$ & $  4.57\pm00.44$ & $  7.35\pm00.45$\\
$-22.25<M<-22.00$ & $  2.43\pm02.42$ & $  0.97\pm00.43$ & $  0.69\pm00.26$ & $  1.49\pm00.30$ & $  3.06\pm00.36$ & $  5.52\pm00.39$\\
$-22.50<M<-22.25$ & $  0.69\pm00.69$ & $  1.43\pm00.52$ & $  0.31\pm00.17$ & $  0.50\pm00.17$ & $  0.94\pm00.20$ & $  2.52\pm00.26$\\
$-22.75<M<-22.50$ & $  0.67\pm00.66$ & $  0.78\pm00.39$ & $  0.29\pm00.17$ & $  0.68\pm00.20$ & $  0.64\pm00.17$ & $  0.99\pm00.16$\\
$-23.00<M<-22.75$ & $  0.69\pm00.67$ & $  0.54\pm00.47$ & \nodata & \nodata & $  0.18\pm00.09$ & $  0.73\pm00.15$
\enddata
\tablenotetext{a}{$M=M_{^{0.1}r}-5 {\rm log} h$}
\end{deluxetable}

\begin{deluxetable}{crrrrrr}

\tablecolumns{7}
\tablewidth{0pt}
\tabletypesize{\scriptsize}
\tablecaption{AGES $^{0.1}$r-band $1/V_{\rm max}$ Luminosity Functions for Red Galaxies \label{tab:ages_r_Red_lf}}
\tablehead{
\colhead{} & 
\multicolumn{6}{c}{Luminosity Function ($\times 10^{-4}$ $h^3$ Mpc$^{-3}$ mag$^{-1}$)}\\
\colhead{Luminosity Range\tablenotemark{a}} & 
\colhead{$0.05<z<0.15$} & 
\colhead{$0.15<z<0.25$} &
\colhead{$0.25<z<0.35$} &
\colhead{$0.35<z<0.45$} & 
\colhead{$0.45<z<0.55$} & 
\colhead{$0.55<z<0.75$}}
\startdata
$-18.25<M<-18.00$ & $ 17.69\pm03.59$ & \nodata & \nodata & \nodata & \nodata & \nodata\\
$-18.50<M<-18.25$ & $ 20.00\pm03.63$ & \nodata & \nodata & \nodata & \nodata & \nodata\\
$-18.75<M<-18.50$ & $ 28.81\pm04.63$ & $ 23.05\pm02.27$ & \nodata & \nodata & \nodata & \nodata\\
$-19.00<M<-18.75$ & $ 30.45\pm06.64$ & $ 31.94\pm02.48$ & \nodata & \nodata & \nodata & \nodata\\
$-19.25<M<-19.00$ & $ 31.64\pm06.46$ & $ 25.32\pm02.21$ & \nodata & \nodata & \nodata & \nodata\\
$-19.50<M<-19.25$ & $ 37.24\pm04.96$ & $ 33.36\pm02.55$ & \nodata & \nodata & \nodata & \nodata\\
$-19.75<M<-19.50$ & $ 31.26\pm04.55$ & $ 42.45\pm03.13$ & \nodata & \nodata & \nodata & \nodata\\
$-20.00<M<-19.75$ & $ 36.75\pm04.94$ & $ 40.92\pm02.93$ & $ 22.78\pm01.48$ & \nodata & \nodata & \nodata\\
$-20.25<M<-20.00$ & $ 45.19\pm08.76$ & $ 41.77\pm03.08$ & $ 26.98\pm02.00$ & \nodata & \nodata & \nodata\\
$-20.50<M<-20.25$ & $ 25.74\pm04.19$ & $ 34.15\pm02.57$ & $ 21.63\pm01.51$ & \nodata & \nodata & \nodata\\
$-20.75<M<-20.50$ & $ 26.89\pm04.37$ & $ 30.76\pm02.43$ & $ 21.21\pm01.43$ & $ 26.55\pm01.29$ & \nodata & \nodata\\
$-21.00<M<-20.75$ & $ 21.27\pm03.94$ & $ 21.39\pm02.04$ & $ 18.36\pm01.42$ & $ 27.98\pm01.32$ & \nodata & \nodata\\
$-21.25<M<-21.00$ & $ 19.95\pm03.89$ & $ 18.04\pm01.87$ & $ 16.27\pm01.25$ & $ 19.05\pm01.07$ & \nodata & \nodata\\
$-21.50<M<-21.25$ & $ 11.07\pm04.06$ & $ 14.16\pm01.65$ & $ 10.62\pm01.01$ & $ 16.33\pm01.01$ & $ 15.45\pm00.82$ & \nodata\\
$-21.75<M<-21.50$ & $  9.21\pm03.03$ & $  9.35\pm01.34$ & $  5.92\pm00.75$ & $ 11.33\pm00.83$ & $ 14.90\pm00.80$ & \nodata\\
$-22.00<M<-21.75$ & $  5.05\pm02.50$ & $  5.97\pm01.07$ & $  4.54\pm00.66$ & $  5.72\pm00.59$ & $  9.04\pm00.63$ & $  8.16\pm00.47$\\
$-22.25<M<-22.00$ & $  2.34\pm02.34$ & $  2.51\pm00.69$ & $  2.92\pm00.53$ & $  4.70\pm00.53$ & $  6.17\pm00.52$ & $  5.52\pm00.37$\\
$-22.50<M<-22.25$ & $ 25.88\pm25.06$ & $  1.02\pm00.44$ & $  1.35\pm00.36$ & $  1.93\pm00.34$ & $  4.29\pm00.43$ & $  6.56\pm00.40$\\
$-22.75<M<-22.50$ & \nodata & $  0.61\pm00.35$ & $  0.48\pm00.21$ & $  1.46\pm00.30$ & $  1.76\pm00.28$ & $  2.53\pm00.24$\\
$-23.00<M<-22.75$ & \nodata & \nodata & $  0.19\pm00.14$ & $  0.18\pm00.10$ & $  1.23\pm00.23$ & $  1.49\pm00.18$
\enddata
\tablenotetext{a}{$M=M_{^{0.1}r}-5 {\rm log} h$}
\end{deluxetable}

\begin{deluxetable}{ccccccccc}
\tablecolumns{9}
\tablewidth{0pt}
\tabletypesize{\scriptsize}
\tablecaption{$r^{0.1}$ Luminosity Function Parameters \label{tab:lffits}}
\tablehead{
  \colhead{Galaxy Sample} & 
  \colhead{$z$} & 
 \colhead{$M_*$} & 
   \colhead{error} & 
  \colhead{$\alpha$} & 
  \colhead{$\phi_*$} & 
  \colhead{\% error} & 
  \colhead{$j$} & 
  \colhead{$\sigma_{\mbox{LSS}}$/$j$} \\
 & & & & & \colhead{$h^3 {\rm Mpc}^{-3} {\rm Mag}^{-1}$} & & \colhead{$10^8 hL_\sun{\rm Mpc}^{-3}$} & }
  \startdata
All Galaxies & 0.10 & $-20.58 $& 0.05 & $-1.05$ & 0.0159 & 11 &$ 2.25 \pm  0.65 $& 0.28 \\
... & 0.20 & $-20.81 $& 0.04 & $-1.05$ & 0.0152 &  5 &$ 2.67 \pm  0.56 $& 0.20 \\
... & 0.30 & $-20.81 $& 0.03 & $-1.05$ & 0.0124 &  7 &$ 2.16 \pm  0.39 $& 0.17 \\
... & 0.40 & $-20.99 $& 0.04 & $-1.05$ & 0.0144 &  7 &$ 2.97 \pm  0.45 $& 0.14 \\
... & 0.50 & $-21.29 $& 0.08 & $-1.05$ & 0.0108 & 10 &$ 2.93 \pm  0.41 $& 0.13 \\
... & 0.65 & $-21.38 $& 0.06 & $-1.05$ & 0.0105 & 14 &$ 3.09 \pm  0.36 $& 0.08 \\
Blue Galaxies & 0.10 &$ -20.31$ & 0.12&$-1.11 $&0.0111 & 16 &$ 1.27 \pm  0.36$ & 0.28\\
... & 0.20 &$ -20.57$ & 0.07&$-1.11 $&0.0101 &  6 &$ 1.49 \pm  0.31$ & 0.20\\
... & 0.30 &$ -20.54$ & 0.04&$-1.11 $&0.0091 &  9 &$ 1.29 \pm  0.23$ & 0.17\\
... & 0.40 &$ -20.78$ & 0.05&$-1.11 $&0.0090 &  9 &$ 1.59 \pm  0.24$ & 0.14\\
... & 0.50 &$ -20.93$ & 0.11&$-1.11 $&0.0081 & 15 &$ 1.65 \pm  0.23$ & 0.13\\
... & 0.65 &$ -21.13$ & 0.08&$-1.11 $&0.0089 & 14 &$ 2.17 \pm  0.31$ & 0.08\\
Red Galaxies & 0.10 &$ -20.49$ & 0.07 & $-0.55$ & 0.0084 & 15 & $0.93 \pm  0.29 $& 0.28\\
... & 0.20 &$ -20.63$ & 0.04 & $-0.55$ & 0.0089 &  8 & $1.12 \pm  0.25 $& 0.20\\
... & 0.30 &$ -20.74$ & 0.04 & $-0.55$ & 0.0060 & 10 & $0.84 \pm  0.17 $& 0.17\\
... & 0.40 &$ -20.84$ & 0.04 & $-0.55$ & 0.0082 &  8 & $1.26 \pm  0.20 $& 0.14\\
... & 0.50 &$ -21.19$ & 0.07 & $-0.55$ & 0.0060 & 12 & $1.28 \pm  0.20 $& 0.13\\
... & 0.65 &$ -21.46$ & 0.06 & $-0.55$ & 0.0036 & 14 & $0.97 \pm  0.31 $& 0.30\\
\enddata
\end{deluxetable}

\begin{deluxetable}{cccc}
\tablecolumns{4}
\tablewidth{0pt}
\tabletypesize{\scriptsize}
\tablecaption{$r^{0.1}$-band Evolution Parameters\label{tab:lf_rparam}}
\tablehead{
  \colhead{Galaxy Sample} & 
  \colhead{$Q$} &
  \colhead{$P$} & 
  \colhead{$n$} \\
  \colhead{} & 
  \colhead{$M_*(z)=M_*(0)+Qz$} & 
  \colhead{$\phi_*(z)\propto1+Pz$} & 
  \colhead{$\j_{^{0.1}r}(z)\propto(1+z)^n$}}
\startdata
All Galaxies & $-1.67\pm0.07$ & $-0.59\pm0.14$ &  $0.81\pm0.27$ \\
Blue Galaxies & $-1.66\pm0.09$ & $-0.38\pm0.21$ & $1.64\pm0.39$ \\
Red Galaxies & $-1.73\pm0.07$ & $-0.95\pm0.10$ &  $0.54\pm0.64$ 
\enddata
\end{deluxetable}

\begin{deluxetable}{crrrrrr}

\tablecolumns{7}
\tablewidth{0pt}
\tabletypesize{\scriptsize}
\tablecaption{AGES $B$-band $1/V_{\rm max}$ Luminosity Functions for All Galaxies \label{tab:ages_B_All_lf}}
\tablehead{
\colhead{} & 
\multicolumn{6}{c}{Luminosity Function ($\times 10^{-4}$ $h^3$ Mpc$^{-3}$ mag$^{-1}$)}\\
\colhead{Luminosity Range\tablenotemark{a}} & 
\colhead{$0.05<z<0.15$} & 
\colhead{$0.15<z<0.25$} &
\colhead{$0.25<z<0.35$} &
\colhead{$0.35<z<0.45$} & 
\colhead{$0.45<z<0.55$} & 
\colhead{$0.55<z<0.75$}}
\startdata
$-18.25<M<-18.00$ & $107.74\pm20.73$ & $107.44\pm19.41$ & \nodata & \nodata & \nodata & \nodata\\
$-18.50<M<-18.25$ & $100.15\pm19.06$ & $106.17\pm16.21$ & \nodata & \nodata & \nodata & \nodata\\
$-18.75<M<-18.50$ & $ 71.43\pm17.53$ & $111.57\pm13.54$ & \nodata & \nodata & \nodata & \nodata\\
$-19.00<M<-18.75$ & $ 62.89\pm16.12$ & $ 74.71\pm11.31$ & \nodata & \nodata & \nodata & \nodata\\
$-19.25<M<-19.00$ & $ 52.47\pm14.82$ & $ 59.86\pm09.45$ & \nodata & \nodata & \nodata & \nodata\\
$-19.50<M<-19.25$ & $ 60.86\pm13.63$ & $ 67.22\pm07.89$ & $ 73.83\pm07.59$ & \nodata & \nodata & \nodata\\
$-19.75<M<-19.50$ & $ 41.44\pm12.53$ & $ 52.24\pm06.59$ & $ 57.96\pm05.77$ & \nodata & \nodata & \nodata\\
$-20.00<M<-19.75$ & $ 26.88\pm11.52$ & $ 34.54\pm05.51$ & $ 46.85\pm04.39$ & \nodata & \nodata & \nodata\\
$-20.25<M<-20.00$ & $ 22.89\pm10.59$ & $ 27.06\pm04.60$ & $ 34.76\pm03.34$ & $ 35.38\pm04.47$ & \nodata & \nodata\\
$-20.50<M<-20.25$ & $ 25.35\pm09.74$ & $ 17.44\pm03.84$ & $ 23.18\pm02.54$ & $ 27.31\pm03.19$ & \nodata & \nodata\\
$-20.75<M<-20.50$ & $  5.68\pm08.96$ & $ 13.12\pm03.21$ & $ 16.41\pm01.93$ & $ 17.96\pm02.27$ & \nodata & \nodata\\
$-21.00<M<-20.75$ & $  3.72\pm08.24$ & $  8.99\pm02.68$ & $  9.80\pm01.47$ & $ 11.14\pm01.62$ & $ 14.63\pm01.91$ & \nodata\\
$-21.25<M<-21.00$ & $  7.40\pm07.57$ & $  4.52\pm02.24$ & $  4.61\pm01.12$ & $  6.45\pm01.16$ & $  9.64\pm01.31$ & \nodata\\
$-21.50<M<-21.25$ & $  4.03\pm06.96$ & $  1.63\pm01.87$ & $  2.69\pm00.85$ & $  3.43\pm00.83$ & $  5.17\pm00.89$ & $  9.45\pm00.67$\\
$-21.75<M<-21.50$ & $  0.10\pm06.40$ & $  0.48\pm01.56$ & $  0.73\pm00.65$ & $  1.94\pm00.59$ & $  2.77\pm00.61$ & $  5.25\pm00.50$\\
$-22.00<M<-21.75$ & \nodata & \nodata & $  0.38\pm00.49$ & $  0.54\pm00.42$ & $  0.86\pm00.42$ & $  2.00\pm00.37$\\
$-22.25<M<-22.00$ & $  1.98\pm05.41$ & $  0.67\pm01.09$ & \nodata & $  0.31\pm00.30$ & $  0.31\pm00.28$ & $  1.10\pm00.28$
\enddata
\tablenotetext{a}{$M=M_{B}-5 {\rm log} h$}
\end{deluxetable}

\begin{deluxetable}{crrrrrr}

\tablecolumns{7}
\tablewidth{0pt}
\tabletypesize{\scriptsize}
\tablecaption{AGES $B$-band $1/V_{\rm max}$ Luminosity Functions for Red Galaxies \label{tab:ages_B_Red_lf}}
\tablehead{
\colhead{} & 
\multicolumn{6}{c}{Luminosity Function ($\times 10^{-4}$ $h^3$ Mpc$^{-3}$ mag$^{-1}$)}\\
\colhead{Luminosity Range\tablenotemark{a}} & 
\colhead{$0.05<z<0.15$} & 
\colhead{$0.15<z<0.25$} &
\colhead{$0.25<z<0.35$} &
\colhead{$0.35<z<0.45$} & 
\colhead{$0.45<z<0.55$} & 
\colhead{$0.55<z<0.75$}}
\startdata
$-18.25<M<-18.00$ & $ 79.30\pm13.68$ & $ 82.83\pm09.09$ & \nodata & \nodata & \nodata & \nodata\\
$-18.50<M<-18.25$ & $ 75.34\pm13.87$ & $ 71.20\pm07.79$ & \nodata & \nodata & \nodata & \nodata\\
$-18.75<M<-18.50$ & $ 47.95\pm14.05$ & $ 82.05\pm06.68$ & \nodata & \nodata & \nodata & \nodata\\
$-19.00<M<-18.75$ & $ 35.01\pm14.24$ & $ 41.23\pm05.72$ & \nodata & \nodata & \nodata & \nodata\\
$-19.25<M<-19.00$ & $ 29.74\pm14.44$ & $ 29.90\pm04.90$ & \nodata & \nodata & \nodata & \nodata\\
$-19.50<M<-19.25$ & $ 36.16\pm14.63$ & $ 33.14\pm04.20$ & $ 42.71\pm04.13$ & \nodata & \nodata & \nodata\\
$-19.75<M<-19.50$ & $ 15.53\pm14.83$ & $ 23.60\pm03.60$ & $ 32.19\pm03.20$ & \nodata & \nodata & \nodata\\
$-20.00<M<-19.75$ & $ 11.74\pm15.03$ & $ 10.31\pm03.09$ & $ 22.35\pm02.47$ & \nodata & \nodata & \nodata\\
$-20.25<M<-20.00$ & $  6.70\pm12.07$ & $  4.48\pm02.65$ & $ 14.69\pm01.91$ & $ 14.81\pm02.43$ & \nodata & \nodata\\
$-20.50<M<-20.25$ & $ 13.09\pm15.43$ & $  2.38\pm02.27$ & $  8.06\pm01.48$ & $  9.31\pm01.79$ & \nodata & \nodata\\
$-20.75<M<-20.50$ & $  0.57\pm01.02$ & $  2.88\pm01.94$ & $  4.56\pm01.14$ & $  6.28\pm01.31$ & \nodata & \nodata\\
$-21.00<M<-20.75$ & $  0.37\pm00.67$ & $  2.52\pm01.67$ & $  3.12\pm00.88$ & $  2.97\pm00.97$ & $  7.46\pm01.78$ & \nodata\\
$-21.25<M<-21.00$ & $  3.21\pm05.78$ & $  0.89\pm01.43$ & $  1.29\pm00.68$ & $  1.72\pm00.71$ & $  5.02\pm01.19$ & \nodata\\
$-21.50<M<-21.25$ & $  4.03\pm07.25$ & $  1.43\pm01.22$ & $  1.03\pm00.53$ & $  1.30\pm00.52$ & $  2.71\pm00.79$ & $  4.21\pm00.67$\\
$-21.75<M<-21.50$ & $  0.01\pm00.02$ & $  0.05\pm00.09$ & $  0.44\pm00.41$ & $  1.55\pm00.39$ & $  1.67\pm00.53$ & $  2.45\pm00.48$\\
$-22.00<M<-21.75$ & \nodata & \nodata & $  0.23\pm00.32$ & $  0.48\pm00.28$ & $  0.79\pm00.35$ & $  1.33\pm00.35$\\
$-22.25<M<-22.00$ & $  0.20\pm00.36$ & $  0.67\pm00.77$ & \nodata & $  0.15\pm00.21$ & $  0.31\pm00.23$ & $  0.77\pm00.25$
\enddata
\tablenotetext{a}{$M=M_{B}-5 {\rm log} h$}
\end{deluxetable}

\begin{deluxetable}{crrrrrr}

\tablecolumns{7}
\tablewidth{0pt}
\tabletypesize{\scriptsize}
\tablecaption{AGES $B$-band $1/V_{\rm max}$ Luminosity Functions for Blue Galaxies \label{tab:ages_B_Blue_lf}}
\tablehead{
\colhead{} & 
\multicolumn{6}{c}{Luminosity Function ($\times 10^{-4}$ $h^3$ Mpc$^{-3}$ mag$^{-1}$)}\\
\colhead{Luminosity Range\tablenotemark{a}} & 
\colhead{$0.05<z<0.15$} & 
\colhead{$0.15<z<0.25$} &
\colhead{$0.25<z<0.35$} &
\colhead{$0.35<z<0.45$} & 
\colhead{$0.45<z<0.55$} & 
\colhead{$0.55<z<0.75$}}
\startdata
$-18.25<M<-18.00$ & $ 28.44\pm07.85$ & $ 24.62\pm09.22$ & \nodata & \nodata & \nodata & \nodata\\
$-18.50<M<-18.25$ & $ 24.81\pm07.96$ & $ 34.97\pm07.90$ & \nodata & \nodata & \nodata & \nodata\\
$-18.75<M<-18.50$ & $ 23.48\pm08.07$ & $ 29.52\pm06.77$ & \nodata & \nodata & \nodata & \nodata\\
$-19.00<M<-18.75$ & $ 27.88\pm08.17$ & $ 33.48\pm05.80$ & \nodata & \nodata & \nodata & \nodata\\
$-19.25<M<-19.00$ & $ 22.73\pm08.28$ & $ 29.96\pm04.97$ & \nodata & \nodata & \nodata & \nodata\\
$-19.50<M<-19.25$ & $ 24.69\pm08.40$ & $ 34.08\pm04.26$ & $ 31.11\pm04.83$ & \nodata & \nodata & \nodata\\
$-19.75<M<-19.50$ & $ 25.90\pm08.51$ & $ 28.64\pm03.65$ & $ 25.77\pm03.74$ & \nodata & \nodata & \nodata\\
$-20.00<M<-19.75$ & $ 15.14\pm08.62$ & $ 24.22\pm03.13$ & $ 24.50\pm02.89$ & \nodata & \nodata & \nodata\\
$-20.25<M<-20.00$ & $ 16.18\pm08.74$ & $ 22.59\pm02.68$ & $ 20.08\pm02.23$ & $ 20.56\pm02.86$ & \nodata & \nodata\\
$-20.50<M<-20.25$ & $ 12.26\pm08.86$ & $ 15.06\pm02.30$ & $ 15.12\pm01.73$ & $ 18.00\pm02.10$ & \nodata & \nodata\\
$-20.75<M<-20.50$ & $  5.11\pm08.98$ & $ 10.24\pm01.97$ & $ 11.85\pm01.33$ & $ 11.69\pm01.55$ & \nodata & \nodata\\
$-21.00<M<-20.75$ & $  3.34\pm09.10$ & $  6.46\pm01.69$ & $  6.68\pm01.03$ & $  8.17\pm01.14$ & $  7.18\pm01.45$ & \nodata\\
$-21.25<M<-21.00$ & $  4.19\pm09.22$ & $  3.63\pm01.45$ & $  3.33\pm00.80$ & $  4.73\pm00.84$ & $  4.62\pm00.97$ & \nodata\\
$-21.50<M<-21.25$ & \nodata & $  0.21\pm01.24$ & $  1.67\pm00.62$ & $  2.14\pm00.62$ & $  2.46\pm00.64$ & $  6.28\pm00.57$\\
$-21.75<M<-21.50$ & $  0.09\pm09.47$ & $  0.43\pm01.06$ & $  0.29\pm00.48$ & $  0.39\pm00.45$ & $  1.10\pm00.43$ & $  3.35\pm00.41$\\
$-22.00<M<-21.75$ & \nodata & \nodata & $  0.15\pm00.37$ & $  0.05\pm00.33$ & $  0.07\pm00.29$ & $  0.81\pm00.29$\\
$-22.25<M<-22.00$ & $  1.78\pm09.73$ & \nodata & \nodata & $  0.16\pm00.25$ & $  0.01\pm00.19$ & $  0.39\pm00.21$
\enddata
\tablenotetext{a}{$M=M_{B}-5 {\rm log} h$}
\end{deluxetable}

\begin{deluxetable}{ccccccccc}
\tablecolumns{9}
\tablewidth{0pt}
\tabletypesize{\scriptsize}
\tablecaption{$B$-band Luminosity Function Parameters \label{tab:lf_bfits}}
\tablehead{
  \colhead{Galaxy Sample} & 
  \colhead{$z$} & 
 \colhead{$M_*$} & 
   \colhead{error} & 
  \colhead{$\alpha$} & 
  \colhead{$\phi_*$} & 
  \colhead{\% error} & 
  \colhead{$j$} & 
  \colhead{$\sigma_{\mbox{LSS}}$/$j$} \\
 & & & & & \colhead{$10^{-3} h^3 {\rm Mpc}^{-3} {\rm Mag}^{-1}$} & & \colhead{$10^8 hL_\sun{\rm Mpc}^{-3}$} & }
  \startdata
All Galaxies & 0.10 & -19.92 & 0.05 & -1.20 &  8.41 & 10 & $1.58\pm0.50$ & 0.28 \\
... & 0.20 & -20.04 & 0.04 & -1.20 &  8.97 &  4 & $1.88\pm0.46$ & 0.20 \\
... & 0.30 & -20.05 & 0.04 & -1.20 & 10.84 &  8 & $2.30\pm0.34$ & 0.17 \\
... & 0.40 & -20.25 & 0.04 & -1.20 &  8.69 &  7 & $2.22\pm0.38$ & 0.14 \\
... & 0.50 & -20.44 & 0.08 & -1.20 &  7.95 & 10 & $2.40\pm0.35$ & 0.13 \\
Blue Galaxies & 0.10 & -19.65 & 0.12 & -1.30 &  6.03 & 16 & $0.89\pm0.32$ & 0.28 \\
... & 0.20 & -19.77 & 0.09 & -1.30 &  6.52 &  5 & $1.07\pm0.29$ & 0.20 \\
... & 0.30 & -19.90 & 0.04 & -1.30 &  5.74 &  8 & $1.06\pm0.23$ & 0.17 \\
... & 0.40 & -20.05 & 0.07 & -1.30 &  6.73 & 10 & $1.42\pm0.23$ & 0.14 \\
... & 0.50 & -20.26 & 0.13 & -1.30 &  6.47 & 14 & $1.66\pm0.21$ & 0.13 \\
Red Galaxies & 0.10 & -19.63 & 0.09 & -0.50 &  6.94 & 16 & $0.68\pm0.17$ & 0.28 \\
... & 0.20 & -19.67 & 0.06 & -0.50 &  8.42 &  7 & $0.86\pm0.16$ & 0.20 \\
... & 0.30 & -19.72 & 0.05 & -0.50 &  7.91 & 10 & $0.85\pm0.11$ & 0.17 \\
... & 0.40 & -19.81 & 0.06 & -0.50 &  7.44 &  8 & $0.86\pm0.12$ & 0.14 \\
... & 0.50 & -19.89 & 0.08 & -0.50 &  6.09 & 10 & $0.76\pm0.13$ & 0.13 \\
\enddata
\end{deluxetable}

\begin{deluxetable}{ccccc}
\tablecolumns{5}
\tablewidth{0pt}
\tabletypesize{\scriptsize}
\tablecaption{AGES Galaxy Selection Criteria and Sparse Sampling Rates \label{tab:selection}}
\tablehead{
  \colhead{Sample Name} & 
  \colhead{gshort Bit} & 
  \colhead{Bright Sample Limits} & 
  \colhead{Faint Sample Limits} & 
  \colhead{Faint Sample Sparse Sampling Rate}}
\startdata
Main $I$-band Sample & 4096 & $15.45<I<18.95$ & $18.95<I<20.45$ & 20\% \\
$R$-band Sample & 1024 & $R\le19.41$ & $19.41\le R\le20.21$ & 20\% \\
$B_w$-band Sample & 512 & $B_w\le20.5$ & $20.5\le B_w\le21.3$ & 20\% \\
$J$-band Sample\tablenotemark{a} & 256 & $J\le 18.42$ & $18.42\le J\le19.42$ & 20\% \\
$K_s$-band Sample\tablenotemark{b} & 128 & $K_s\le17.84$ & $17.84\le K_s\le18.34$ & 20\% \\
GALEX NUV Sample & 64 & $NUV<21$ & $21<NUV<22$ & 30\% \\ 
GALEX FUV Sample & 32 & $FUV<22$ & $22<FUV<22.5$ & 30\% \\
3.6$\mu$m Sample & 16 & $[3.6\mu m]<18$ &  $18\le[3.6\mu m]\le18.5$ & 30\% \\
4.5$\mu$m Sample & 8 & $[4.5\mu m]<18.46$ &  $18.46\le[4.5\mu m]\le18.96$ & 30\% \\
5.8$\mu$m Sample & 4  & $[5.8\mu m]<18.43$ &  $18.43\le[5.8\mu m]\le18.93$ & 30\% \\
8.0$\mu$m sample & 2 & $[8.0\mu m]<18.2$ &  $18.2\le[8.0\mu m]\le18.8$ & 30\% \\
MIPS 24$\mu$m Sample & 1  & $F_{24}\ge 0.5$ mJy & $0.3{\rm mJy} \le F_{24} < 0.5 {\rm mJy}$ & 30\% \\
\enddata
\tablenotetext{a}{$J$-band photometry used here comes entirely from FLAMEX}
\tablenotetext{b}{$K_s$-band photometry came both from FLAMEX and NWDFS imaging.  In constructing the $K_s$ sample,
if photometry from either survey met our criteria, it was included in the sample}.
\end{deluxetable}
\begin{deluxetable}{ccrrrc}
	
	\tablecolumns{6}
	\tablewidth{0pt}
	\tabletypesize{\scriptsize}
	\tablecaption{AGES Main Galaxy Subsample Statistics\label{tab:samplesize}}
	\tablehead{
		\colhead{Sample Name} &
		\colhead{Sampling Rate} &  
		\colhead{Targets} &
		\colhead{Spectra} & 
		\colhead{Redshifts} & 
		\colhead{Completeness}
	}
\startdata
MIPS & 30\% & 4662 & 4484 & 4411 & 95\% \\
IRAC [8.0] & 30\% & 3536 & 3498 & 3490 & 99\% \\
IRAC [5.8] & 30\% & 4058 & 3982 & 3927 & 98\% \\
IRAC [4.5] & 30\% & 6215 & 6081 & 5999 & 98\% \\
IRAC [3.6] & 30\% & 4992 & 4882 & 4792 & 98\% \\
GALEX FUV & 30\% & 545 & 422 & 520 & 96\% \\
GALEX NUV & 30\% & 1836 & 1779 & 1775 & 97\% \\
K-band & 20\% & 5399 & 5314 & 5302 & 98\% \\
J-band & 20\% & 4319 & 4288 & 4278 & 98\% \\
B-band & 20\% & 4345 & 4278 & 4237 & 99\% \\
R-band & 20\% & 7480 & 7378 & 7304 & 99\% \\
Other I-band & ... & 18368 & 8257 & 7727 & 42\% \\
Main I-band & 20\% &  11011 & 10640 & 10306 & 94\% 
\enddata
\end{deluxetable}

\end{document}